\documentclass[11pt,preprint]{aastex62}
\shorttitle{Systematic Serendipity}
\shortauthors{Giles \& Walkowicz}

\usepackage{listings} % For code
\usepackage{graphicx} % For figures
\usepackage{amsmath} % For some math

\begin{document}

\title{Systematic Serendipity: A Test of Unsupervised Machine Learning as a Method for Anomaly Detection}
\author{Daniel Giles}\affil{Astronomy Department, The Adler Planetarium, Chicago, IL 60605}\affil{Illinois Institute of Technology}\affil{LSSTC Data Science Fellow}
\author{Lucianne Walkowicz}\affil{Astronomy Department, The Adler Planetarium, Chicago, IL 60605}
\email{LWalkowicz@adlerplanetarium.org}
\email{dgiles1@hawk.iit.edu}

%% Mark off your abstract in the ``abstract'' environment. In the manuscript
%% style, abstract will output a Received/Accepted line after the
%% title and affiliation information. No date will appear since the author
%% does not have this information. The dates will be filled in by the
%% editorial office after submission.

\begin{abstract}
Advances in astronomy are often driven by serendipitous discoveries. As survey astronomy continues to grow, the size and complexity of astronomical databases will increase, and the ability of astronomers to manually scour data and make such discoveries decreases. In this work, we introduce a machine learning-based method to identify anomalies in large datasets to facilitate such discoveries, and apply this method to long cadence lightcurves from NASA's Kepler Mission. Our method clusters data based on density, identifying anomalies as data that lie outside of dense regions. This work serves as a proof-of-concept case study and we test our method on four quarters of the Kepler long cadence lightcurves. We use Kepler's most notorious anomaly, Boyajian's Star (KIC 8462852), as a rare `ground truth' for testing outlier identification to verify that objects of genuine scientific interest are included among the identified anomalies. We evaluate the method's ability to identify known anomalies by identifying unusual behavior in Boyajian's Star, we report the full list of identified anomalies for these quarters, and present a sample subset of identified outliers that includes unusual phenomena, objects that are rare in the Kepler field, and data artifacts. By identifying $<$4\% of each quarter as outlying data, we demonstrate that this anomaly detection method can create a more targeted approach in searching for rare and novel phenomena.
\end{abstract}


\section{Introduction}\label{sec_intro}
Survey astronomy is producing more data than ever before, both expanding the number of objects observed and the number of observations per object.
PanSTARRS, for example, recently delivered to astronomy the first petabyte scale data release \citep{Chambers2016}, Gaia has released data for nearly 2 billion sources \citep{Gaia2016,Gaia2018}, and others, like the Transiting Exoplanet Survey Satellite \citep[TESS;][]{TESS}, and the Zwicky Transient Facility \citep[ZTF;][]{ZTF}, have launched and will release data in short order. The Large Synoptic Survey Telescope \citep[LSST;][]{Lsst2009} will have first light in the next few years and deliver 10 to 30 terabytes of data per night. These surveys yield unprecedented insights into the universe by observing billions of stars and galaxies through space and time, adding new objects to every category of known phenomena, and creating new categories of previously unknown, unobserved events.
Identifying new, anomalous, and outlying observations pose a significant challenge given the scale of data. In this work we present a proof-of-concept for a methodology we've developed to address this challenge.

As Douglas Hawkins puts it,``An outlier is an observation which deviates so much from the other observations as to arouse suspicions that it was generated by a different mechanism'' \citep{Hawkins1980}. The need for, and by extension the application of, anomaly detection in large scale astronomy is still relatively new, but anomaly detection is well precedented outside of the astronomical community. Computer scientists have developed techniques to identify abnormalities for a multitude of reasons, including detecting network attacks \citep{Agrawal2015}, fraud \citep{Ahmed2016278}, and malware \citep{Menahem2009}. A survey of different anomaly detection methods is presented by \citet{Chandola2009}. In astronomy, discoveries of novel phenomena have often been more serendipitous than intentional (see \cite{Thompson2012,Boyajian2016,Wright2014}). The scale of modern astronomical surveys does not lend itself to discoveries of anomalies by happenstance, rather there must be a concerted effort to mine the data with machine-based methods to have any hope of identifying anomalous, or outlying data.

Broadly speaking, machine learning falls into two categories: supervised and unsupervised learning, which largely relate to the goals of classifying and clustering, respectively \citep{Ivezic2013}.
In supervised classification, data are sorted into predetermined categories that must be taught to the algorithm through well studied training sets. This method is well suited to quickly identifying objects of known categories with sizable training sets, but is poorly suited to finding rare, novel, or anomalous objects. Finding these sorts of objects after classification generally require a concerted effort to scour the data. Unsupervised clustering, on the other hand, groups data based on a cluster metric (i.e. proximity or density) in feature space. Unsupervised methods do not require a training set, nor do they require initial categories to create groupings. Likewise, these methods do not carry the implication or requirement of prior knowledge pertaining to any underlying mechanisms driving the measured features. Notably, those mechanisms likely exist (e.g. lightcurves of Cepheid variables are self-similar because they share an underlying physical cause), and can potentially be discovered as a result of studying clustered data. Anomalies in clustered data are apparent in the form of outlying data, i.e. data that is unclustered.

The strength of unsupervised learning not requiring training data, however, also carries the issue of validation. Particularly in attempting to successfully identify outlying data, there is no simple, universal way to establish a ground truth as anomalies can only be defined in relation to other data. Anomaly detection in astronomy seeks outliers that have the nebulous quality of being scientifically interesting. In this work, we take a look at a single object known to exhibit aberrant behavior that is of scientific interest: KID 8462852, also known as Boyajian's star for the first author on its discovery paper. Boyajian's star was serendipitously discovered to have unusual behaviors by citizen scientists working on Planet Hunters, a Zooniverse project to identify transiting exoplanets in the Kepler data \citep{Boyajian2016}. The discrepant behavior identified by citizen scientists, asymmetrical dips of varying duration at non-periodic times, can be seen in Quarters 8 and 16 of the Kepler data, shown in the second and fourth panels of Figure \ref{lc_tabby}.  Boyajian's star also has an observed longterm dimming trend \citep{Montet2016,Meng2017} but is otherwise a typical, main-sequence F3 V star (T\textsubscript{eff}=6584K, log g=4.124, mass=1.4M\textsubscript{\(\odot\)}, radius=1.699R\textsubscript{\(\odot\)}). 
It caught public attention for its odd behavior, and interest was further fueled by the potential explanation of this behavior being due to an alien megastructure \citep{Wright2014}. It has been the subject of intense public and scientific scrutiny, garnering 44 references on ADS between the discovery of its behavior in 2016 at the time of this writing. 
As of January 2018, its erratic behavior has been most consistent with an occulter of ordinary dust \citep{Boyajian2018}.

In this paper, we propose a method for identifying potential outliers in astronomical databases, and use the behavior of the well-known, occasionally anomalous source Boyajian's star as a proof-of-concept by evaluating its identifications in different quarters after applying the methodology described below.
In the next section we discuss the properties of the data we consider; in Section \ref{sec_methods} we describe our methods; in Section \ref{sec_results} we present our results on anomaly detection, in which we highlight a small sample of identified outliers including Boyajian's star; and in Section \ref{sec_conclusion} we discuss future directions and applications of our work. 

\section{Data}\label{sec_keplerdata}
The data we consider in this study are long-cadence photometric lightcurves from Quarters 4, 8, 11, and 16 of NASA's {\it Kepler} mission. We utilize Data Release 25 which reprocessed all Q0-Q17 data with the updated data pipeline. We summarize some key features of the Kepler mission and the data we utilize here, but full specifications for Kepler can be found for instrumentation  \citep{KeplerBook}, data characteristics \citep{KeplerDataChar}, data processing \citep{KeplerDataProc}, and the input catalog \citep{Batalha2010}.\footnote{ \footnotesize These resources are available via the Mikulski Archive for Space Telescopes at http://archive.stsci.edu/kepler/}

The Kepler spacecraft was designed to obtain near-continuous photometry for stars in a single, star-rich 105 deg$^2$ field of view (FOV) centered at R.A. = 19h22m40s and Dec = 44$^{\circ}$30'00'' from March 2009 to May 2013. The photometer camera contains 42 CCDs with 2200$\times$1024 pixels, where each pixel covers 4 arcsec. However, only pre-selected stars of interest were downloaded \citep{Batalha2010}. The primary goal of the mission was to identify the fraction of terrestrial exoplanets located in the habitable zone of their host star. The Kepler Mission took incredibly well-sampled and precise observations, achieving about 30ppm for solar type stars \citep{Gilliland2011,Gilliland2015}. Stars where an exoplanetary transit signature of around 100ppm is impossible to detect (i.e., giants, stars fainter than 16th mag, stars in overcrowded fields) were omitted from the target list. Of the roughly half-million targets in the FOV brighter than 16th mag, approximately 30\% were targeted. Beyond the primary target list, additional high priority targets included all known eclipsing binaries in the FOV ($>$600), all members of open clusters in the FOV, and the nearest main sequence stars.
For this work we have utilized the long-cadence observations which are composed of 270, 6.02s exposures totaling about a half hour per observation and over 4,000 observations per epoch. Four times a year, every 3 months, the Kepler spacecraft rolled by 90 deg to re-align its solar panels, and these define epochs known as ``Quarters.'' This will place any given star in one of four different positions on the focal plane depending on season, in this study Quarters 4, 8, and 16 are the same orientation with Quarter 11 in the preceding orientation.

The calibration pipeline for Kepler lightcurves is optimized toward the goal of identifying exoplanetary transits; lightcurves for a particular target are not necessarily free from artifacts. The primary means to identify and clean instrumental signatures and systematic errors, the Presearch Data Conditioning pipeline, corrects or removes affected data where possible, but does not perform well for systematics that are non-temporally correlated between targets.
Details on the Kepler data processing are available in the Kepler Data Processing Handbook \citep{KeplerDataProc}, and known, ongoing phenomena are documented in the Kepler Data Characteristics Handbook \citep{KeplerDataChar}. We do not attempt to remove any remaining artifacts in the Kepler data prior to our analysis, and expect that those artifacts may show up as anomalies (indeed, the identification of artifacts is one of the motivations for this work-- in an ongoing survey, such as TESS, the ability to identify artifacts may result in changes to observing that improve mission data quality overall). 

We demonstrate our method for this case study on four quarters of data: two that feature Boyajian's star exhibiting no noteworthy activity or variability (quarters 4 and 11), and two that feature Boyajian's star exhibiting the unusual behaviour that has attracted the attention and speculation of astronomers world wide (quarters 8 and 16). The data from each quarter is pared down to only the sources that appear in all four quarters to facilitate comparison between quarters.

\section{Methods}\label{sec_methods}
In this section, we describe our method for identifying anomalous objects in Kepler data. We begin by processing photometric data into numerical features, detail how we cluster data based on the derived features, and finally discuss how we evaluate outlier identifications.\footnote{\footnotesize  Our code is made publicly available at https://github.com/d-giles/KeplerML} 

We utilize the standard python machine learning package, scikit-learn \citep{scikit-learn}, and  Astropy, a community-developed core Python package for Astronomy \citep{astropy}, for the purposes of our analysis. We also make heavy use of  SciPy packages 
\citep{scipy} including NumPy \citep{numpy}, Matplotlib \citep{matplotlib}, and Pandas \citep{pandas}.

\subsection{Feature Calculation} \label{sec_features}
The methods here described have been developed for application to the Kepler data, but are applicable to time-series data in general.
The data from the Kepler mission are consistently well-sampled and of regular duration, however, astronomical data in general are not. In the interest of eventually applying these methods to sparser data, we treat Kepler data in the same way we would other data. Instead of clustering based on the photometric data itself, we derive a set of 60 numerical features which describe the lightcurve.
These features were created as part of previous data mining work on the Kepler lightcurves \citep{Walkowicz2014AAS},
several of which are drawn from the features prescribed in \citet{Richards2011} and others developed specifically for Kepler analysis. The features developed and evaluated by \citet{Richards2011} emphasize utility in separating classes of known phenomena and were found to substantially facilitate this. Where features were deemed less important, their inclusion or removal had minimal impact on the results of classification. Initial principle component analyses indicated that more than 40 of the features were required to explain 90\% of the variance and we opted to include all features used in the previous Kepler work. The full set of features, and a brief description of each, can be found in Appendix \ref{app_features}.
Using derived features provides the additional benefit of standardizing clustering compute time after the lightcurves have been initially processed. 
Processing a single, long-cadence lightcurve on a 2.70GHz Intel Xeon CPU running Linux Ubuntu took 6.7s on average. Derived features for each object are saved to a Pandas data frame by quarter. For clustering, all derived features are scaled to have unit variance and shifted to have a zero mean with Scikit-Learn's StandardScaler, but scaled data is not stored.

\subsection{Cluster and Outlier Designation}\label{sec_clustering}
This work utilizes a proximity clustering approach to identify outliers, based on Density-Based Spatial Clustering of Applications with Noise (DBSCAN) \citep{Ester1996}. The DBSCAN algorithm is a nearest neighbor approach with two parameters defining what constitutes a cluster: the maximum separation ($\epsilon$) in feature space between two points to be associated with one another, and the minimum number of associated neighbors ($k$) to qualify a point as a core cluster member. In clustering, DBSCAN will select a random starting point, evaluate whether or not it meets clustering criteria. If it does not, DBSCAN will select a new starting point. Once DBSCAN finds a point that matches the criteria, it examines the neighbors to determine if they match the criteria, if they do DBSCAN evaluates the new points neighbors next, and so on until it finds edge members that neighbor cluster members, but do not enough neighbors within epsilon themselves. Once DBSCAN has identified all members of an individual cluster this way, it will initialize to a new, unexamined point. In this application, we are primarily interested in outliers rather than the clusters themselves. As such, we use the cluster membership definitions from DBSCAN but simplify the approach to return only one of three designations, rather than assigning a cluster label for each object in distinct clusters. The definitions for these designations are laid out in Equation \ref{eq_cluster}. 
\begin{equation}\label{eq_cluster}
Designation[i]=\begin{cases}
		\text{Core Cluster}, & \text{if } d_{i,k}\leq\epsilon\\
        \text{Edge of Cluster}, & \text{if $d_{i,k}>\epsilon$ but any nearer neighbor (j$<$k),}\\
        & \text{$d_{i,j}\leq\epsilon$, is a core cluster member}\\
        \text{Outlier}, & \text{otherwise}
\end{cases}
\end{equation}
All variables in this section are described in Table \ref{table_vars}.

A point may be a core cluster member if it has the minimum number of neighbors within the radius $\epsilon$, an edge cluster member if it contains fewer than $k$ neighbors but has at least one neighbor within $\epsilon$ that is a core cluster member, or an outlier if it does not contain enough neighbors within the cutoff and no neighbors within $\epsilon$ are themselves core cluster members. As density is defined as the number of samples within a given volume, small changes in the $\epsilon$ can produce significantly different clustering results. 

\begin{figure}[tb!]
\gridline{\fig{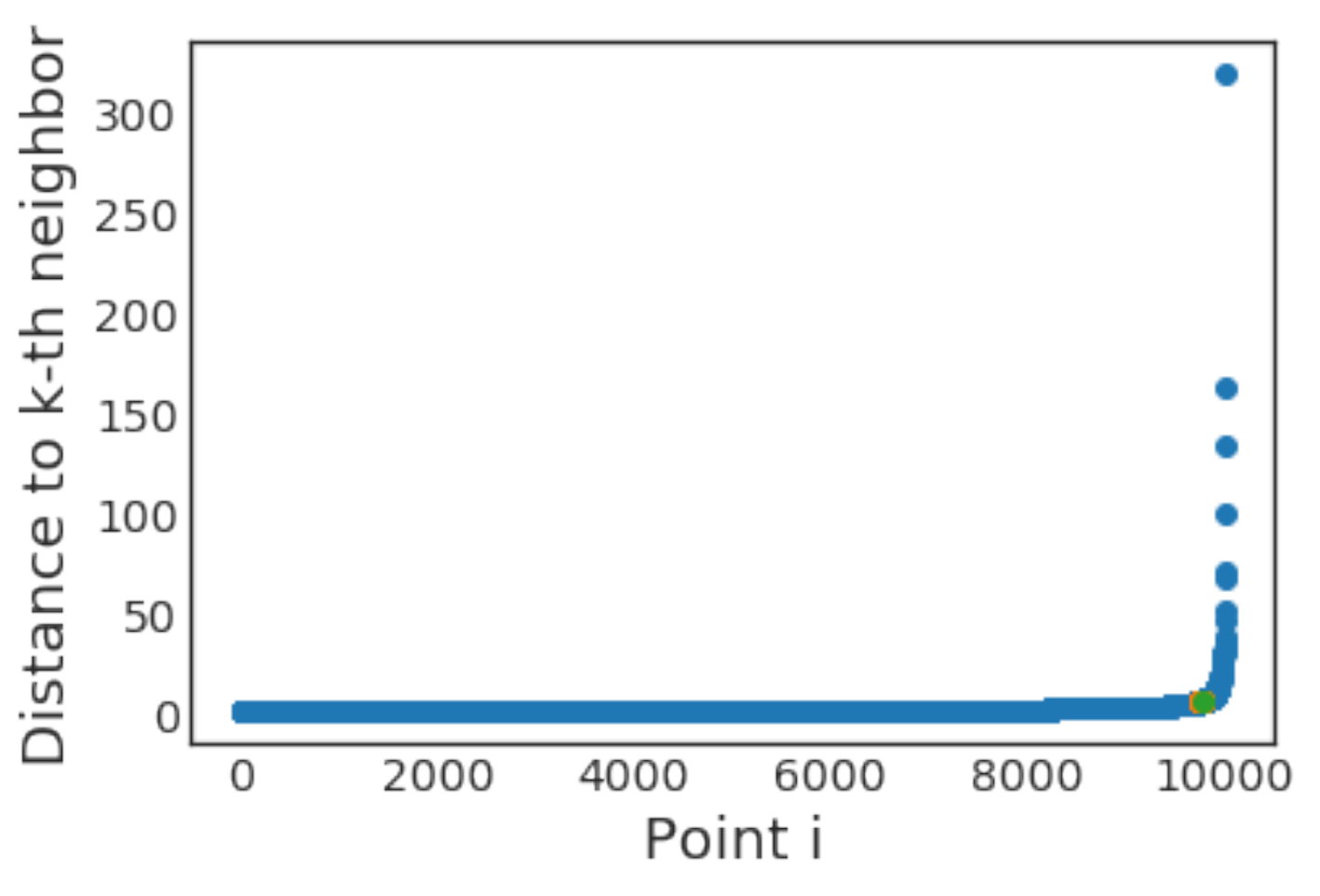}{.5\textwidth}{(a) \footnotesize}
\fig{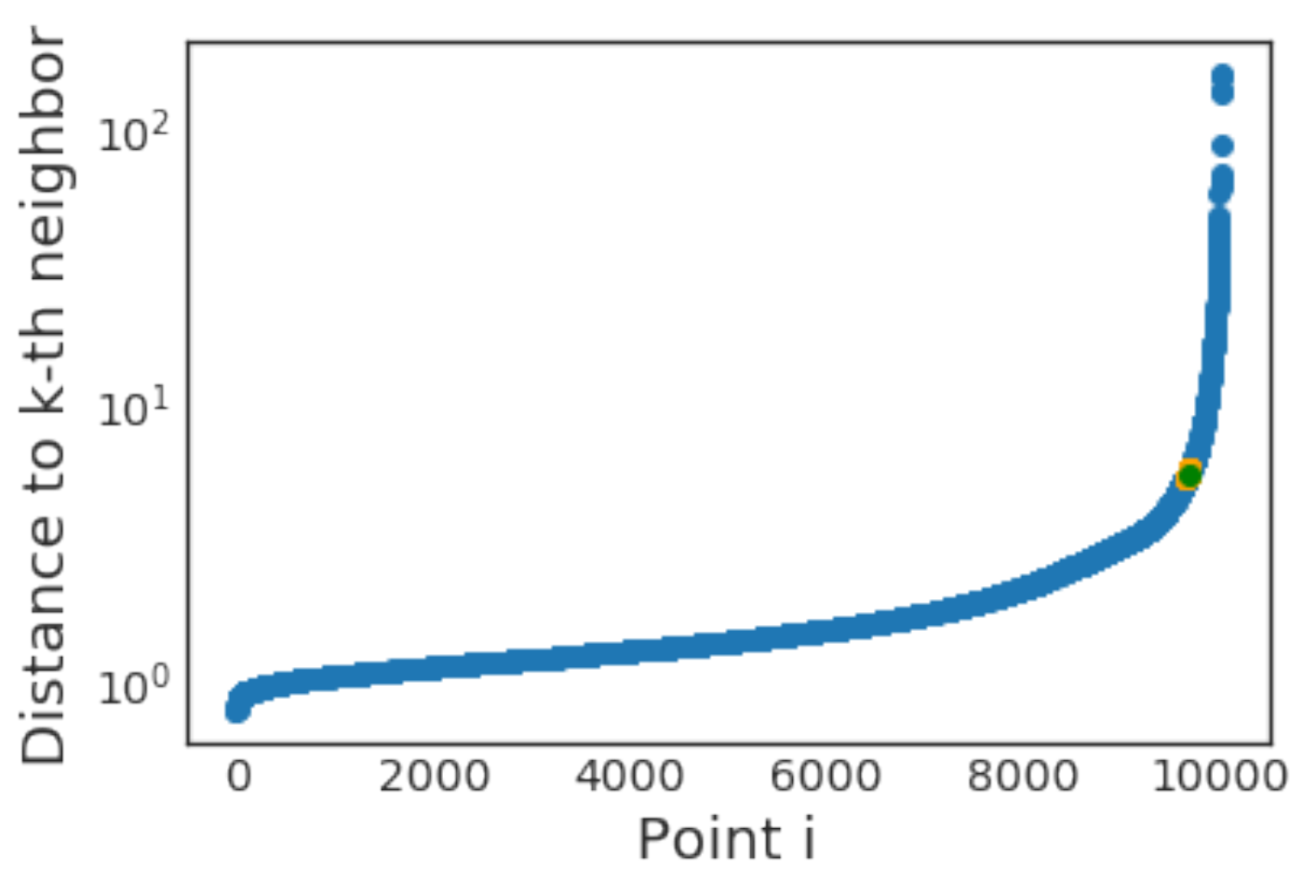}{.5\textwidth}{(b) \footnotesize}
}\caption{\footnotesize A sample of data from Quarter 4. A clear elbow can be seen in the plots illustrating the distance to the $k^{th}$ neighbor. The point in green represents the elbow found by the method described in Section \ref{sec_features}. The points in orange are the points before and after the elbow. (a) shows the entire sample, (b) the elbow in log-scale.}\label{fig_elbow}
\end{figure}

To estimate $\epsilon$ we have adopted the heuristic suggested, and widely utilized, by the original developer of DBSCAN.
The heuristic examines the distance to the $k^{th}$ nearest neighbor for each point, sorts them in order of distance, and finds the elbow where distance to the $k^{th}$ neighbor increases dramatically. The distance at the elbow is defined as $\epsilon$, and the chosen $k$ as the minimum neighbors for a cluster. The elbow occurs where the slope after a point dramatically increases compared with before the point. 
The exact location of an elbow is somewhat subjective as is the resulting epsilon, as can be seen in Figure \ref{fig_elbow} an elbow exists both in linear and log scale, but not at the exact same location. This is somewhat mitigated by the inclusion of edge cluster members which define the edge of a cluster, having a nearby clustered neighbor, but failing to meet cluster criteria. An underestimation of epsilon will identify more core cluster members as edge members, and edge members as outliers. Whereas an overestimation of epsilon will identify more edge members as core cluster members, and more outliers as edge cluster members, in the worst case including all data as core cluster members. In our application of anomaly detection, overlooking anomalous data is the worse offense, and we opt for a conservative epsilon ensuring we catch the early edge of the elbow (as determined in the linear space). 
We automate this heuristic to determine $\epsilon$ for a given sample of data by comparing distances to the $k^{th}$ nearest neighber of values before and after each data point (equivalent to comparing the slopes). For the subset $S$ (defined in Equation \ref{eq_set}) containing all points that match this criteria, $\epsilon$ is defined as the minimum distance of a point to its $k^{th}$ neighbor, Equation \ref{eq_eps}. \\
\begin{equation}\label{eq_set}
S=\{i \mid \sum\limits_{j=i+1}^{i+n}d_{j,k}\geq 1.05 \sum\limits_{j=i-n}^{i-1}d_{j,k}\} 
\end{equation}
\begin{equation}\label{eq_eps}
\epsilon = \min\limits_{i\in S} d_{i,k}
\end{equation}
\begin{equation}\label{eq_k}
k = 4\times\frac{N_{total}}{N_{sample}}%needs to be an integer, figure out the notation
\end{equation}
\citet{Ester1996} indicate that considering $k$ beyond the 4th nearest neighbor has diminishing returns on performance. However, given the size and density of our data after scaling, this approach consistently returns zero for $\epsilon$; the distance is apparently smaller than the precision of our variables can handle. We modify the prescribed heuristic by applying it to a sample of $10^4$ randomly chosen points in order to determine a nonzero $\epsilon$ for the subset, then scale the minimum number of points required for a cluster accordingly.\\
In the data we have considered, which is scaled to unit variance on each feature, distance to the 4th neighbor is consistently flat until the elbow. In the case of the sampled $10^4$ points, we look at 0.2\% of values on either side of each point (20 points). The elbow is determined to be the first point where the average of the following values is at least 5\% greater than the preceding average. This definition is less sensitive to point-to-point variation and is prone to catch the early edge of the elbow as in Figure \ref{fig_elbow}(a). By preempting the elbow a bit and finding lower value of epsilon, cluster membership is more exclusive ensuring data after the true elbow are outliers and more objects on the edge are identified as edge cluster members. We find this preferable to a larger epsilon which would lead to increased core cluster membership and fewer outliers.

\begin{deluxetable*}{Cl}[ht!]
\tablecaption{Cluster input  discussed in Section \ref{sec_clustering} \label{table_vars}}
\tablecolumns{2}
\tablewidth{0pt}
\tablehead{
\colhead{Variable} &
\colhead{Description}
}
\startdata
\epsilon & radius within which neighboring points are considered to be associated\\
k & minimum number of neighbors within $\epsilon$ to qualify a point as part of a cluster\\
i & index of a point in the data\\
j & place-holding integer\\
n & size of range to consider when determining location of elbow\\
d_{i,j} & distance to the $j^{th}$ nearest neighbor of the $i^{th}$ point\\
N_{total} & Size of the entire dataset\\
N_{sample} & Size of the dataset sampled for $\epsilon$ determination\\
S & The set of all potential elbow points, defined in Equation \ref{eq_set}\\
\enddata
\end{deluxetable*}

% \begin{lstlisting}[language=Python]
% for i in d_k[n:-n].index:
% 	pre = np.mean(d_k[i-n:i-1])
%     post = np.mean(d_k[i+1:i+n])
%     if post >= 1.05*pre:
%     	epsilon = d_k[i]
%         break
% \end{lstlisting}\label{code_cutoff}
% Where $\texttt{i}$ is an integer and  $\texttt{d\_k}$ is an array containing the distances to the $k^{th}$ neighbor for each point in the dataset.

\subsection{Dimensionality Reduction and Outlier Evaluation  \label{sec_eval}}
\begin{figure}[htb!] \centering
\includegraphics[width=\textwidth]{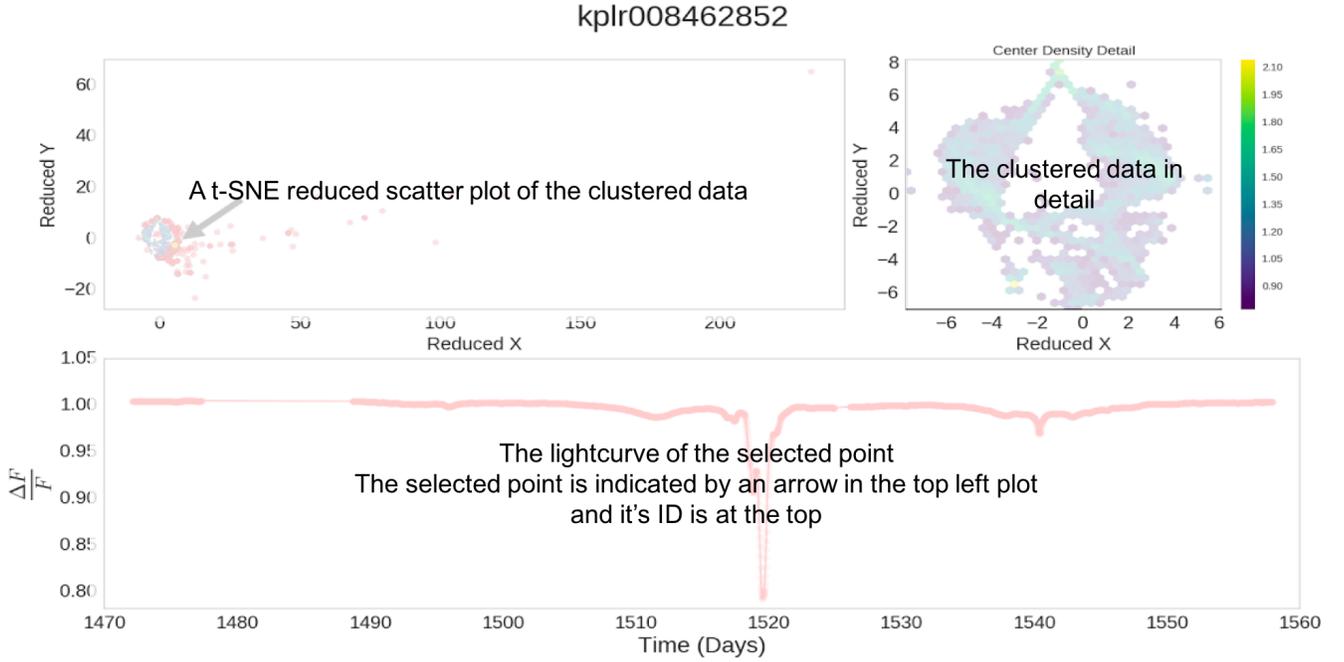}\caption{\footnotesize This GUI facilitates exploration of the data. A user can click on any outlying point in the t-SNE reduced plot and the lightcurve for that object will be displayed in the bottom. Outliers are scattered points in red and clustered data have been log-scale hex-binned into a 35x35 grid to illustrate cluster density. Over 96\% of objects are concentrated in the core cluster.} \label{fig_gui}
\end{figure}

The focus of this initial work is to produce an initial list of outliers and evaluate how this method performs using Boyajian's star as an example. Beyond this, we examine and present a small sample of outlying objects from the full list of outliers produced. We have developed a user interface for visualization and data exploration in Python 2.7. Data exploration relies on a reduction of the data to a 2-dimensional representation that maintains clustering relationships. 
In the GUI, we tie the reduced data to the original lightcurves to explore different clusterings and outliers; selecting a point in the reduction displays that point's associated lightcurve and it's Kepler ID. An example showing Boyajian's star selected in Quarter 16 can be seen in Fig. \ref{fig_gui}.

The clustering method we've used, detailed in Section \ref{sec_clustering}, determines clusters based on density by considering nearest neighbors. In visually representing the high dimensional feature space of the data, we have found the most illustrative reductions focus on maintaining nearest neighbor relationships. One of the best methods for this particular need is t-SNE \citep[t-distributed Stochastic Neighborhood Embedding;][]{VanDerMaaten2008}.
As well as it can, t-SNE preserves the relative proximity of nearest neighbors (via a similarity score) with the consequence of exaggerating the distance of other data points. This is accomplished by calculating similarity scores between points of the full-dimensional set and generating a uniform random distribution in the desired dimensionality assigning each random point to a corresponding data point. The algorithm then calculates the similarity scores for the low dimensional distribution and the differences between the scores for each point, and uses that difference to move each point towards, or away from each other point in the low dimensional space, and this is repeated until the Kullback-Leibler divergence is minimized.
As t-SNE is computationally expensive, it was necessary to break the larger dataset, on the scale of $10^5$, into smaller, $10^4$ scale chunks. 
Unfortunately, the reduction of a sample cannot be simply generalized to a larger population. Since t-SNE fundamentally relies on an n-body simulation, inclusion of additional data would affect the resulting positions of all points.
With that limitation noted, only clustered data has been omitted and subsets of data have proven effective in illustrating clusters and outlier relationships. In following plots, a common sample is used between all four quarters with clustering done on the full quarters. This sample was chosen with two primary considerations: every object identified as an outlier in any quarter are included, plus $10^4$ points are randomly sampled from the remaining objects that were consistently clustered. In each of the full quarters, around 95\% of the objects are clustered, while clustered objects constitute only around two-thirds of each quarter's sample. 

We visually inspect the t-SNE reduced data and cluster determinations. We display their lightcurves to determine if outlier classifications are justified using a GUI developed for exploration as shown in Figure \ref{fig_gui}. We assess the `movement' of individual points in the feature space from quarter to quarter to better understand what type of an outlier each point is as it appears that smaller movements correspond to noise and measurement errors, and larger movements to more drastic feature differences.

\section{Results and Discussion}\label{sec_results}
We discuss the results of our outlier identification method first as a whole, then by quarter. We evaluate how this performed for our proof of concept case, Boyajian's star, and finally we discuss a sample of objects from the outliers including some of the most noticeable outliers from the reduction.

\begin{deluxetable}{cccc}
%% This is the title of the table.
\tablecaption{Outlier Summary}
\tablecolumns{4}
\tablewidth{0pt}
\tablehead{\colhead{Description} & \colhead{Count} & \colhead{Percent of total objects} & \colhead{Percent of Outliers}} 

%% All data must appear between the \startdata and \enddata commands
\startdata
Total Outliers & 8,507 & 5.68\% &  \\
Outliers in every quarter & 3,584 & 2.39\% & 42.13\% \\
Transient Outliers & 4,923 & 3.29\% & 57.87\% \\
\enddata
\end{deluxetable}\label{tab_overall}

\begin{figure}[htb!]\centering
\includegraphics[width=0.45\textwidth]{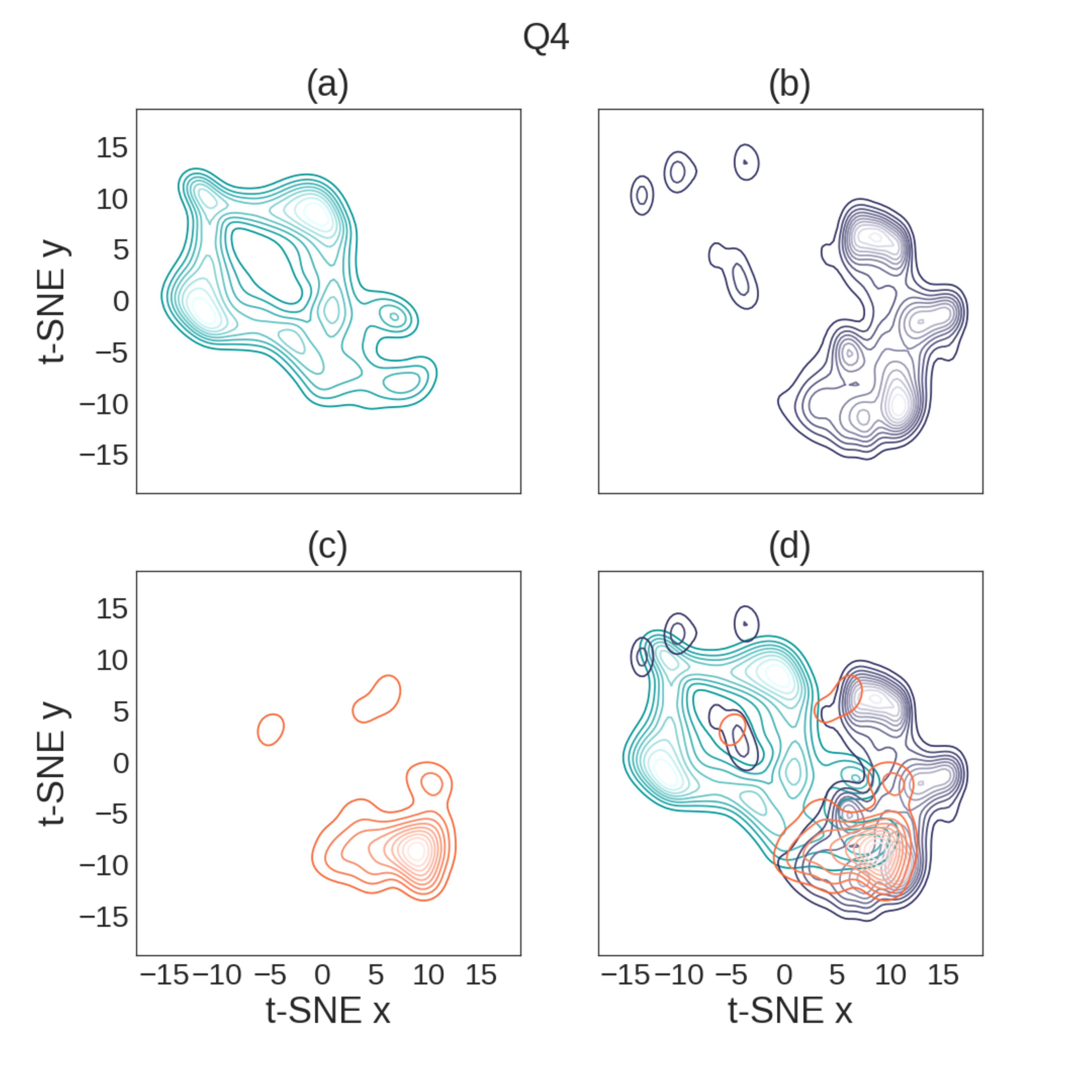}
\includegraphics[width=0.45\textwidth]{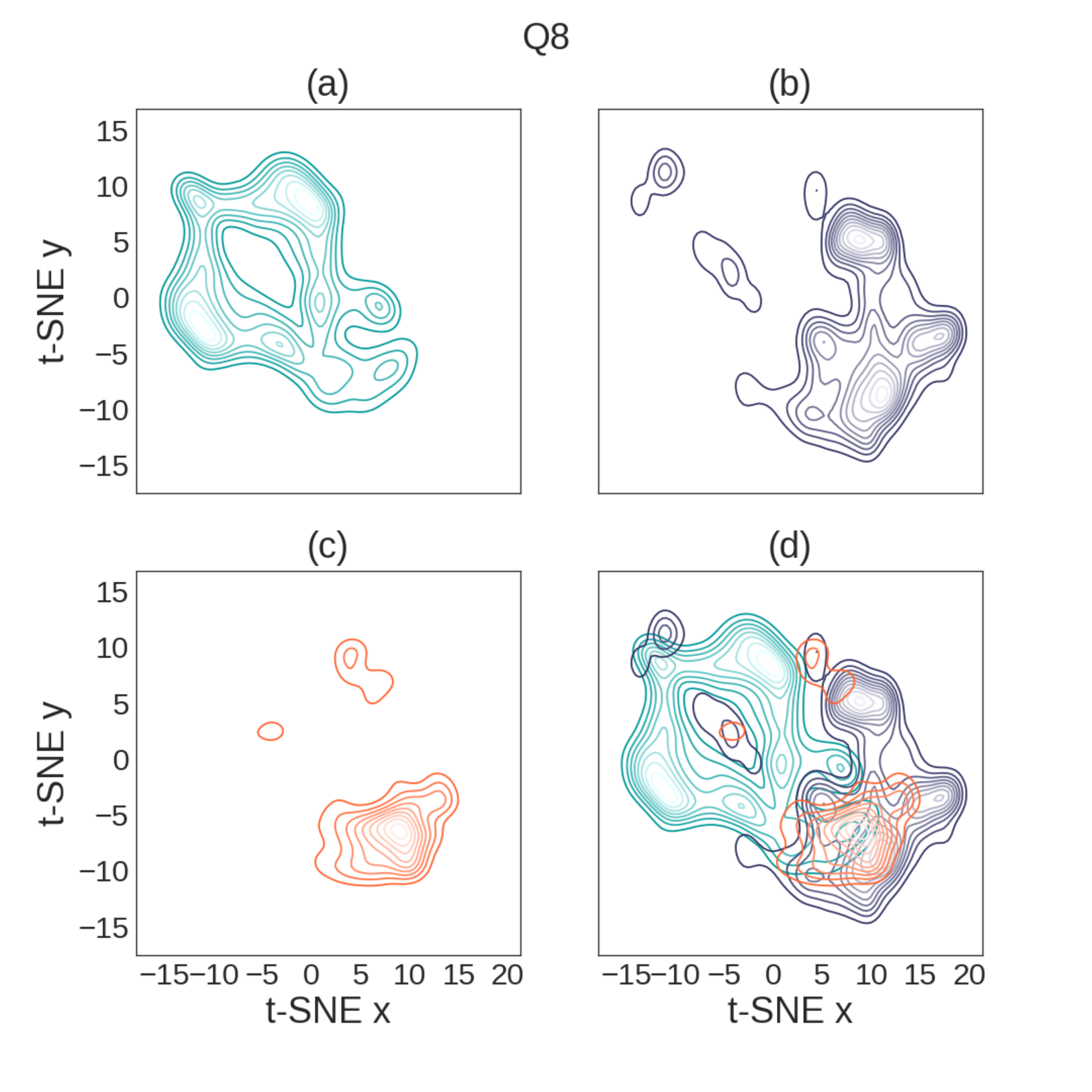}\\
\includegraphics[width=0.45\textwidth]{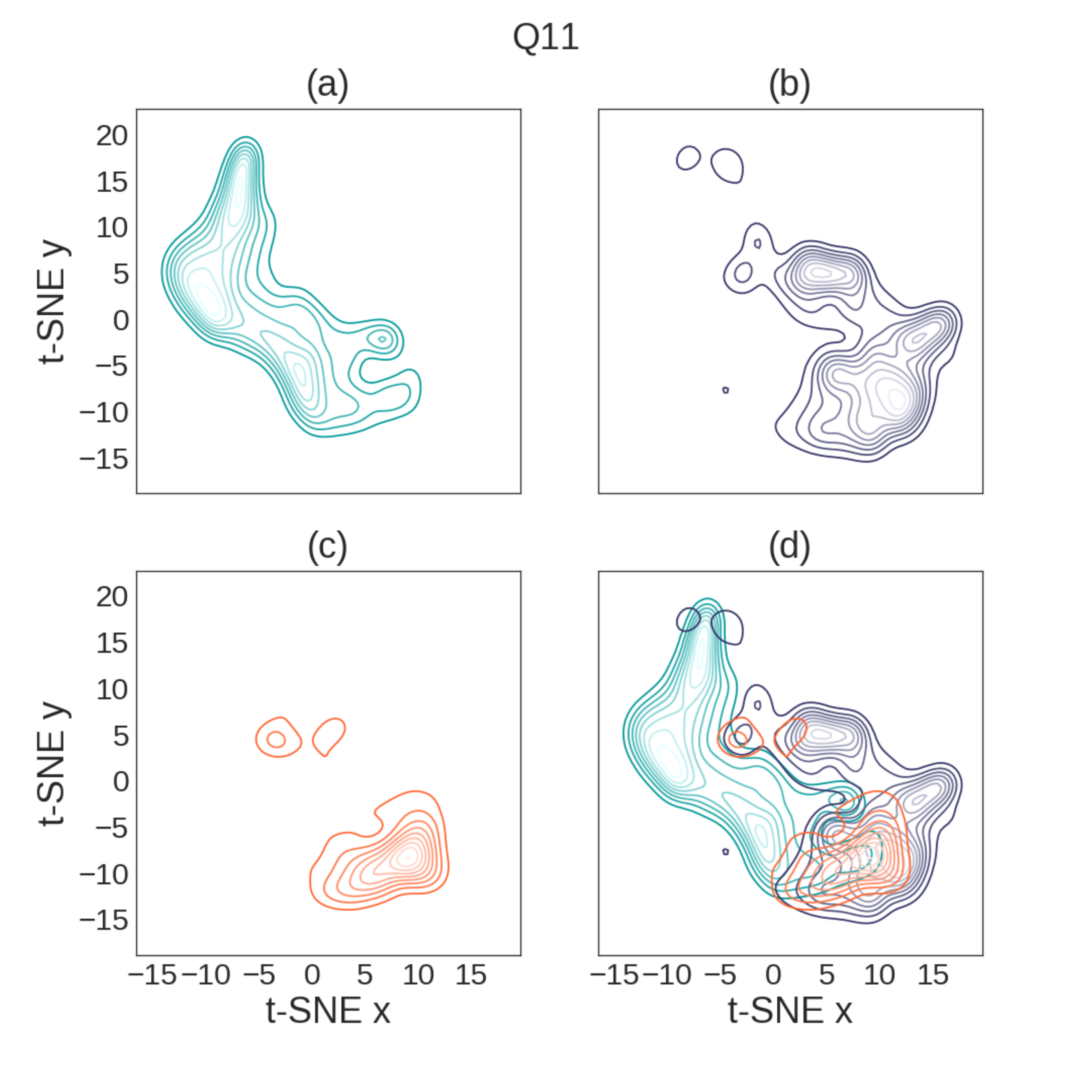}
\includegraphics[width=0.45\textwidth]{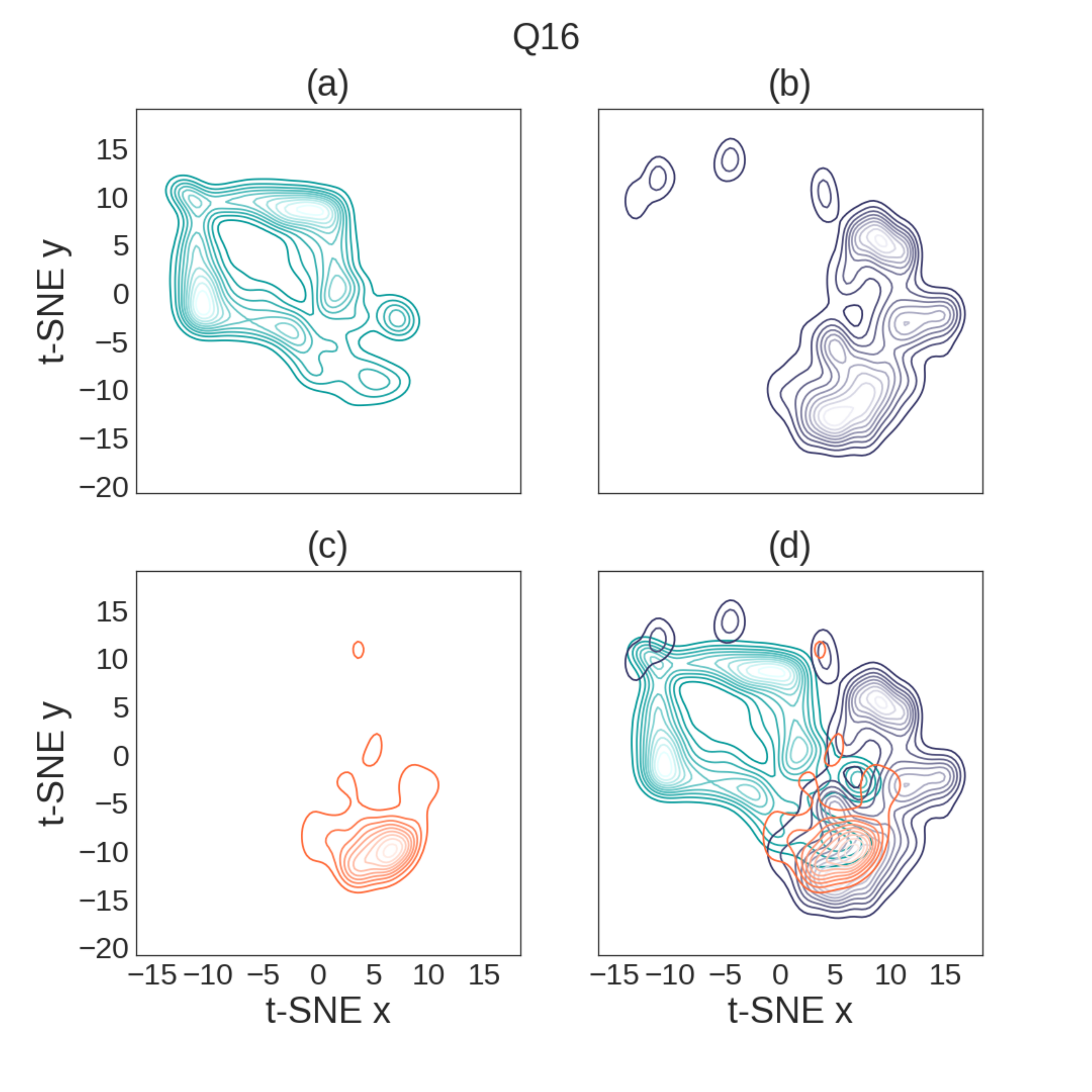}
\caption{\footnotesize Above, we visualize each quarter with a common sample to illustrate how clustered data and the outlying data relates to each other within each quarter. We have chosen the sample such that all objects that are identified as outliers in at least one quarter have been included (8507 objects), alongside 10k additional objects randomly sampled from the remaining data. Outliers constitute up to a third of the data in each quarter's sample, however, in the full Quarter about 96\% of lightcurves are clustered and only only up to 4\% are identified as outliers.
We cluster the full data in 60-D feature space, use t-SNE to represent the sample in 2D, and show the density of data using three Gaussian Kernel Density Estimates for each quarter. In each subplot
(a) shows the KDE for the clustered data of that quarter in blue, (b) shows the KDE for outlying data in purple, (c) shows the KDE for edge-of-cluster data in orange, and (d) shows all three KDEs overplotted. Lighter colored lines indicate higher density regions.}
\label{fig_4panel}
\end{figure}

Across all quarters we considered 149,789 objects, of which 8,507 unique objects were identified as outliers representing 5.68\% of all objects considered. 141,282 objects, 94.32\% of all objects, were identified only as part of a cluster, either as core cluster members or edge cluster members. Objects that were identified as outliers in every quarter constituted 3,584 of the outliers (2.39\% of all objects and 42\% of all outliers), and the remaining 4,923 objects were found to be transient outliers, identified as an outlier and as a cluster member at least once each in different quarters. We summarize the designations made overall in Table \ref{tab_overall} and by quarter in Table \ref{tab_byquarter}. A KDE contour of the t-SNE reduced sample for each quarter is shown in Figure \ref{fig_4panel}.

\begin{deluxetable}{cccc}

%% This is the title of the table.
\tablecaption{Designations By Quarter}

%% The \tablehead gives provides the column headers.
\tablehead{\colhead{Quarter} & \colhead{Core Cluster Members} & \colhead{Edge Cluster Memebers} & \colhead{Outliers} \\ 
\colhead{} & \colhead{(\% of total pop.)} & \colhead{(\% of total pop.)} & \colhead{(\% of total pop.)} } 

%% All data must appear between the \startdata and \enddata commands
\startdata
Q4 & 142,534 & 2,389 & 4,866 \\
 & (95.16\%) & (1.59\%) & (3.25\%) \\
Q8 & 142,235 & 2,510 & 5,044 \\
 & (94.96\%) & (1.68\%) & (3.37\%) \\
Q11 & 140,933 & 2,977 & 5,879 \\
 & (94.09\%) & (1.99\%) & (3.92\%) \\
Q16 & 141,129 & 2,910 & 5,750 \\
 & (94.22\%) & (1.94\%) & (3.84\%) \\
\enddata
\end{deluxetable}\label{tab_byquarter}
We examine the relationships between clustered, edge of cluster, and outlier data in Figure \ref{fig_4panel}. We separate each quarter into the different designations, representing each group with a different Kernel Density Estimate (KDE) plot. While these reductions are limited to a sample only and do not contain the entirety of the data, they are still a helpful visual guide to the data relationships. The general shapes of each KDE plot for each of the designations seem to be fairly consistent, even in Quarter 11 which has the same general form but is lopsided. 

\begin{deluxetable}{rrrrrrrrrrrrrrr}

%% Rotate to a landscape orientation
\rotate
\tablecaption{List of Outliers}

\tablehead{\colhead{KID} & \colhead{RA} & \colhead{Dec} & \colhead{$T_{eff}$} & \colhead{$T_{eff}$ err} & \colhead{$T_{eff}$ err} & \colhead{log g} & \colhead{log g err} & \colhead{log g err} & \colhead{Kepmag} & \colhead{Q4} & \colhead{Q8} & \colhead{Q11} & \colhead{Q16} \\ 
\colhead{} & \colhead{} & \colhead{} & \colhead{(K)} & \colhead{(K)} & \colhead{(K)} & \colhead{} & \colhead{} & \colhead{} & \colhead{} & \colhead{} & \colhead{} & \colhead{} & \colhead{} } 

%% All data must appear between the \startdata and \enddata commands
\startdata
757099 & 19 24 10.334 & +36 35 37.72 & 5519 & 182 & -149 & 3.82 & 0.64 & -0.21 & 13.15 & -1 & -1 & -1 & -1 \\
757450 & 19 24 33.024 & +36 34 38.57 & 5332 & 106 & -96 & 4.50 & 0.05 & -0.04 & 15.26 & 0 & 1 & -1 & 0 \\
892376 & 19 24 15.329 & +36 38 08.92 & 3973 & 124 & -152 & 4.66 & 0.06 & -0.02 & 13.96 & -1 & -1 & -1 & -1 \\
893507 & 19 25 15.007 & +36 37 59.62 & 5382 & 177 & -144 & 3.92 & 0.67 & -0.29 & 12.52 & -1 & 1 & -1 & -1 \\
893647 & 19 25 21.108 & +36 41 24.86 & 4856 & 146 & -131 & 4.58 & 0.06 & -0.04 & 15.28 & -1 & 1 & -1 & -1 \\
1025986 & 19 24 08.086 & +36 46 15.75 & 5604 & 84 & -67 & 4.23 & 0.21 & -0.11 & 10.15 & -1 & 1 & 1 & -1 \\
1026032 & 19 24 10.577 & +36 43 45.38 & 5951 & 160 & -178 & 4.64 & 0.03 & -0.13 & 14.81 & -1 & -1 & -1 & -1 \\
1026474 & 19 24 35.786 & +36 43 25.75 & 4276 & 150 & -165 & 4.60 & 0.05 & -0.02 & 15.28 & 1 & -1 & -1 & -1 \\
1026861 & 19 24 56.220 & +36 43 44.62 & 7063 & 74 & -95 & 4.04 & 0.14 & -0.11 & 11.00 & 0 & -1 & -1 & -1 \\
1026957 & 19 25 01.078 & +36 44 37.00 & 4859 & 97 & -97 & 4.61 & 0.02 & -0.05 & 12.56 & 0 & 0 & 0 & -1 \\
\enddata

\tablecomments{For Q4, Q8, Q11, and Q16: -1 indicates that an object is outlying in that quarter, 0 core cluster membership, and 1 that an object is an edge cluster member. Full, machine readable table available at https://github.com/d-giles/KeplerML/blob/paperwork/KIC\_Outlier\_Properties.csv.}
\end{deluxetable}

\subsection{Results for Boyajian's Star}\label{sec_boyajian}
\begin{figure} \centering
\includegraphics[width=.5\textwidth]{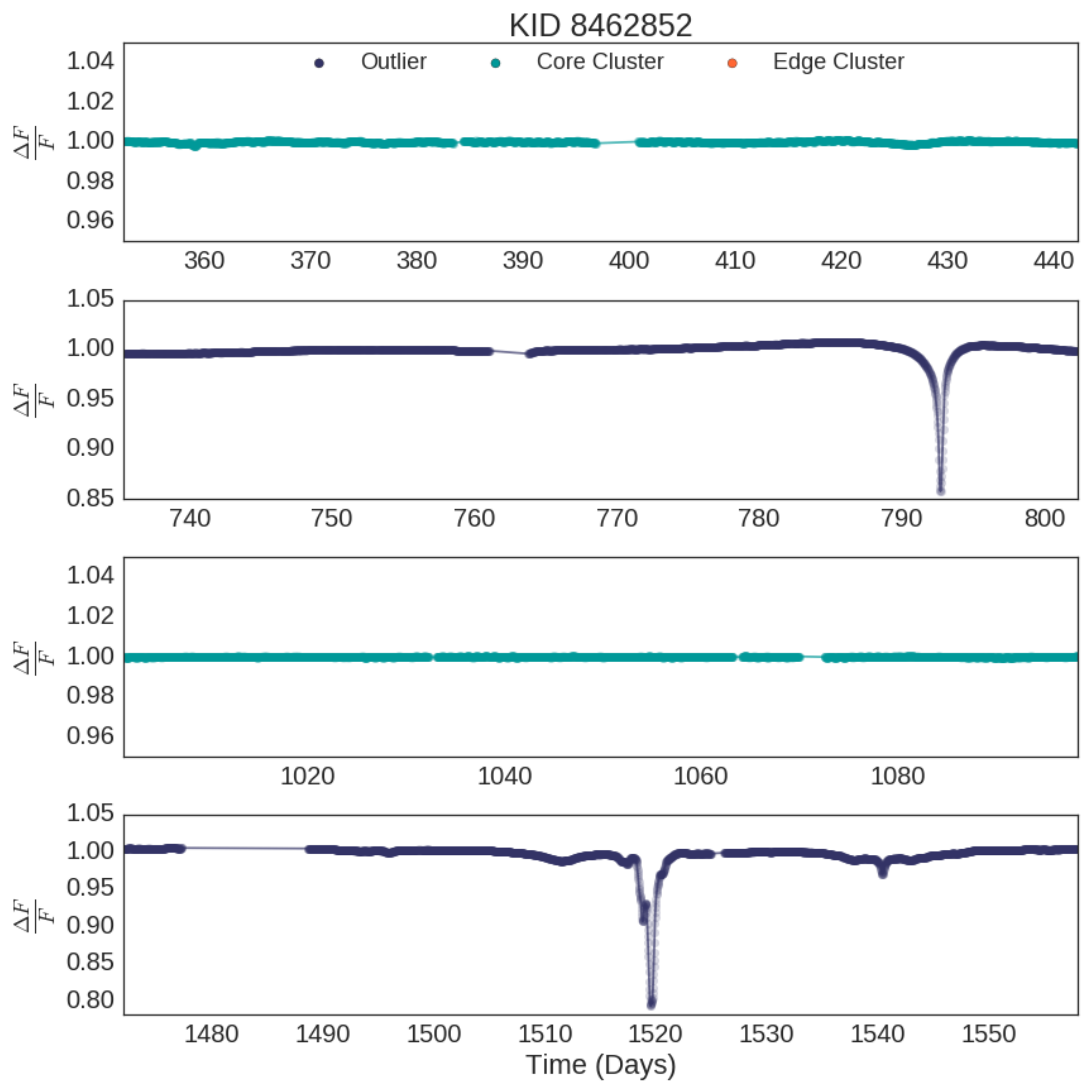}\caption{\footnotesize Here we can see that our method has successfully identified the behaviour of Boyajian's star in each quarter.
Boyajian's star has only been identified as an outlier in  the quarters also noted by scientists and citizen scientists as anomalous \citep{Boyajian2016}.} \label{lc_tabby}
\end{figure}

As mentioned in Section \ref{sec_intro}, we considered Boyajian's star as a proof-of-concept case to evaluate whether or not our method would be able to recreate the serendipitous discovery of anomalous behavior made by human evaluation. As can be seen in Figure \ref{lc_tabby}, our method was able to identify Boyajian's star as an outlier only where odd behavior existed, proving consistent with previous human evaluation. Finding genuine outliers and anomalies of potential interest in hundreds of thousands of sources by circumstance is not trivial, but neither is it impossible (as evidenced by the discovery of Boyajian's star). As we move to millions and billions of sources per survey, however, it becomes much less likely that scientists, or even crowd-sourced searches, will turn up serendipitous discoveries. The method we have developed has identified less than 4\% of the objects as outlying in any given quarter, significantly more manageable than the whole dataset. Coupled with the inclusion of genuine anomalous data like Boyajian's star, this indicates that our methodology can significantly facilitate these discoveries. 
%As the data scales to larger and larger datasets, the volume of clusters will grow and so too will the surface area of those clusters. However, the surface area will grow at a slower rate than the volume and the overall proportion of outliers should decrease, further focusing on the truly anomalous.

In Figure \ref{all_annotated} we look at Boyajian Star's position relative to other data in each quarter. In Quarters 4 and 11 where Boyajian's star exhibits no aberrant behavior, it falls neatly into the core clustered data. In Quarters 8 and 16, Boyajian's star can be found outside the core cluster, illustrating its identification as an outlier.

\subsection{Example Outlier Objects}\label{sec_rand_sample}
\begin{figure}[tb!] \centering
\includegraphics[width=\textwidth]{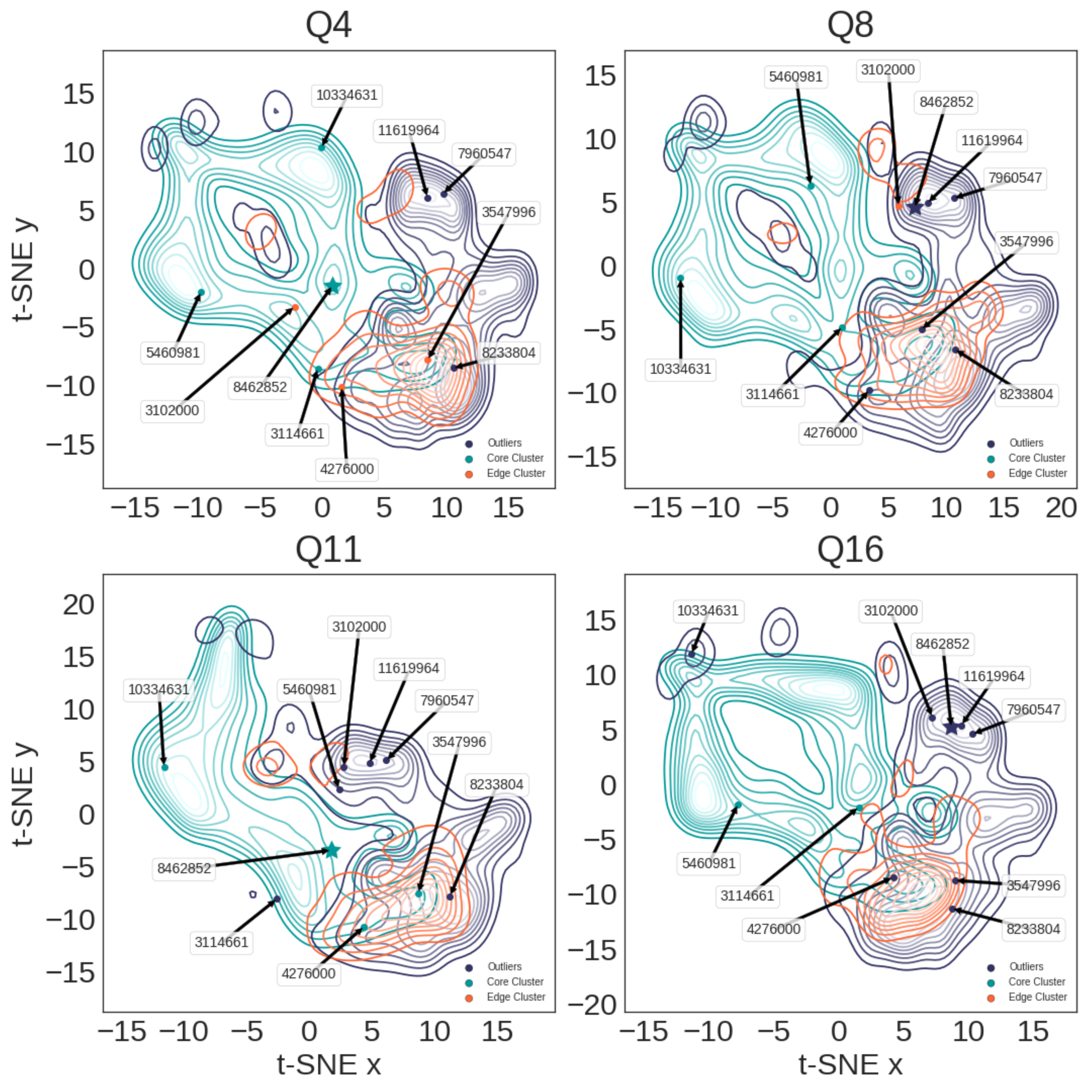}\caption{\footnotesize 
Here we highlight ten objects which have been identified as an outlier in at least one quarter. Three of these objects are outliers in all quarters and the remaining objects are transient outliers that are identified in one or more quarters as an outlier, but not in all quarters. 
Boyajian's star, KID 8462852, itself is a transient outlier in Quarters 8 and 16 but as a core cluster member in Quarters 4 and 11, its lightcurves can be found in Figure \ref{lc_tabby}.}
\label{all_annotated}
\end{figure}

Here we show a few example objects semi-randomly selected from the larger sample of outliers. These examples have been selected to include objects that were identified as outliers in all quarters, some that were identified as an outlier in only one quarter, and others with multiple outlier identifications. Notably, they were not individually selected as specific types of outliers and are presented largely without thorough investigation into the orign of their anomalous behavior. They are provided purely to illustrate a small sample of the outlying data. In Figure \ref{all_annotated} we highlight the example outlier object excepting the most outlying objects discussed in Section \ref{sec_mostoutlying} as their positions in the reduction are significantly removed from the rest of the data. 

\subsubsection{Anomalous Phenomena}\label{sec_mostoutlying}
\begin{figure}
\includegraphics[width=\textwidth]{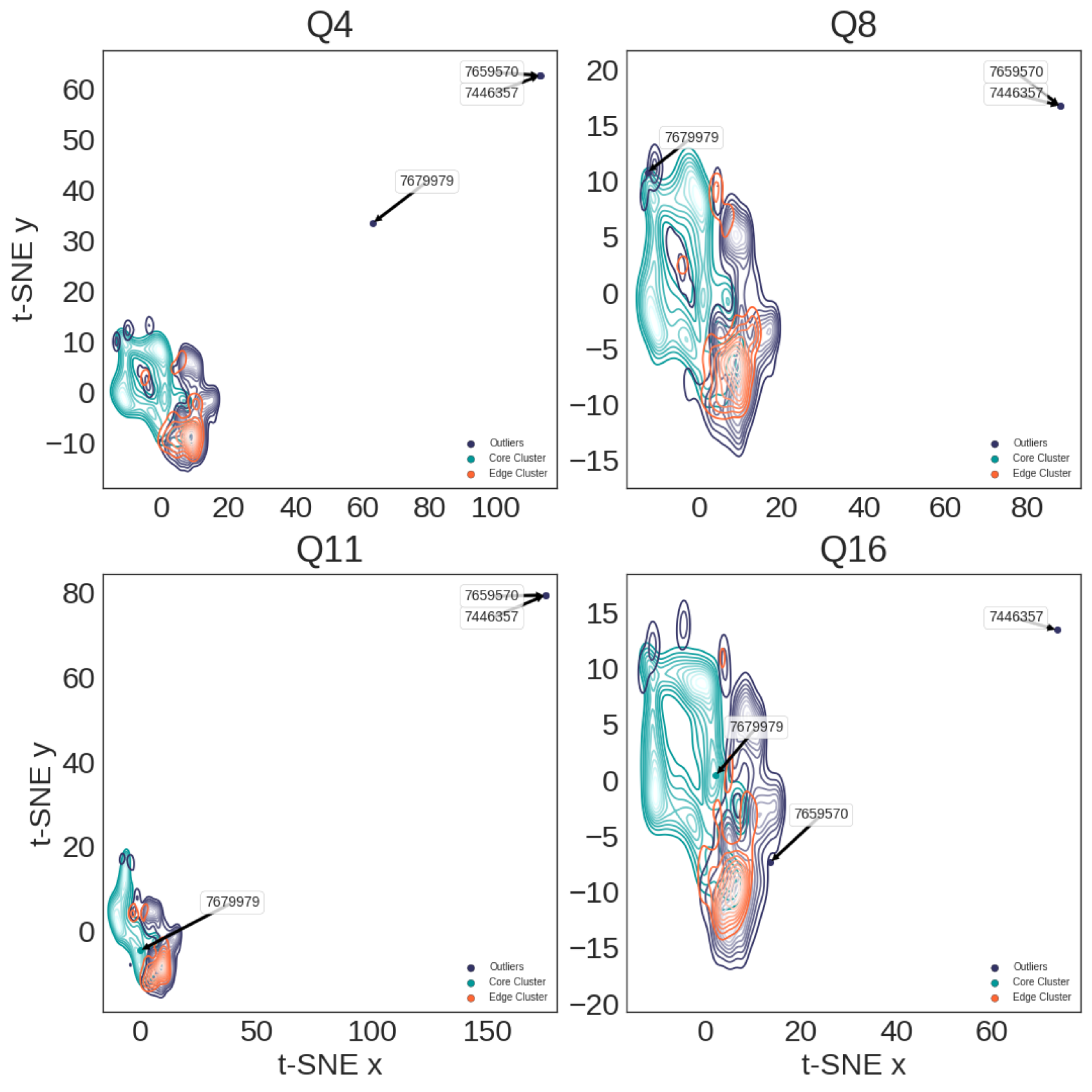}
\caption{\footnotesize Three objects in this sample extend well beyond all other data in the t-SNE reduction. Two of these, KID 7446357 and KID 7659570, are extreme outliers in all quarters and their lightcurves are shown in Figures \ref{lc_SU}(a) and \ref{lcs_rare}(b) respectively. These two objects are SU-Uma cataclysmic variables. KID 7679979 is an extreme outlier only in quarter 4, its lightcurves are shown in Figure \ref{lcs_artifacts}(c) and is included in the discussion of artifacts in Section \ref{artifacts}.} \label{full_scale}
\end{figure}

When we examine the t-SNE reduced data, there are three objects that stand out more than any others: KIC 7679979, KIC 7446357 and KIC 7659570 (Fig. \ref{full_scale}). KIC 7679979 exhibits extreme behavior in only Quarter 4. On inspection of its lightcurve, this does not seem to be due to astrophysical mechanisms, rather it seems more likely to be an artifact and is discussed further in Section \ref{artifacts}.

\begin{figure}[tb!]
\gridline{\fig{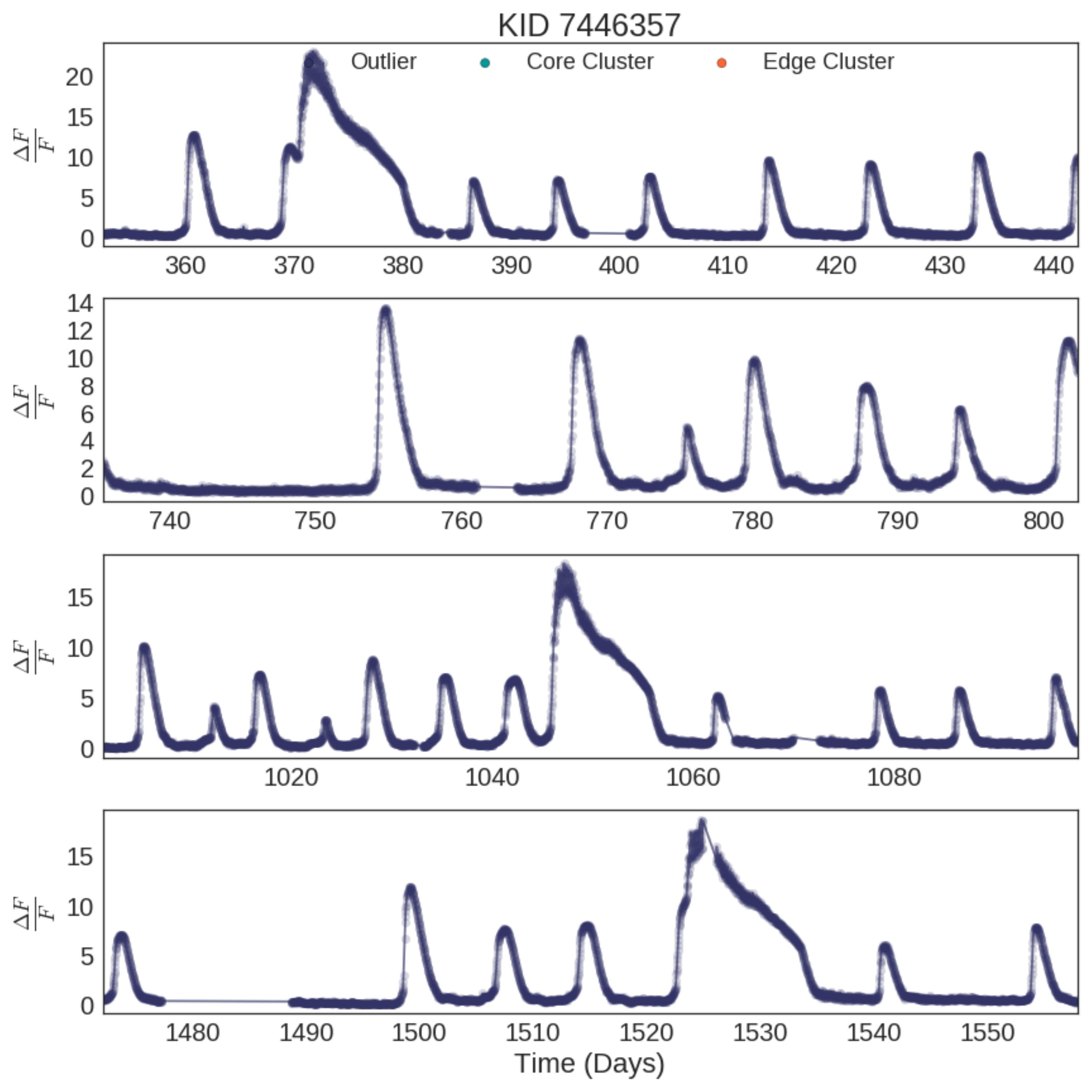}{.5\textwidth}{(a) \footnotesize  KIC 7446357 has been identified as an SU Uma Cataclysmic Variable and as an outlier in each quarter.}
\fig{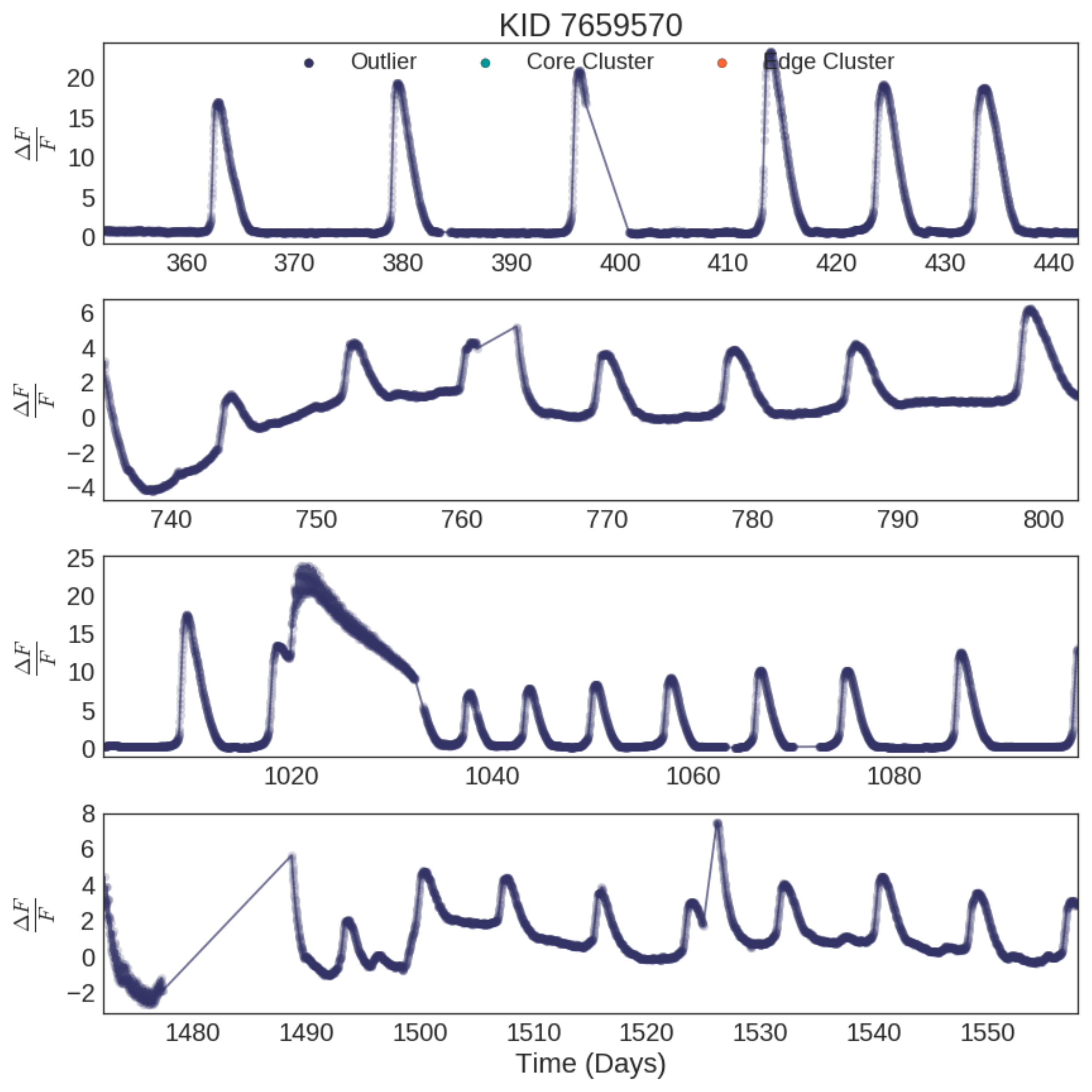}{.5\textwidth}{(b) \footnotesize KIC 7446357 has been identified as an SU Uma Cataclysmic Variable and as an outlier in each quarter. In Quarters 8 and 16, the normalized flux of KIC 7659570 is negative in the first 20 days. These negative fluxes appear to coincide with an outburst at the start of those observation cycles.}
}\caption{\footnotesize SU-Uma Cataclysmic Variables have extreme features, are very rare in the Kepler dataset, and are identified as outliers by our method.} \label{lc_SU}
\end{figure}

KICs 7446357 and 7659570 are V1504 Cyg and V344 Lyr, respectively. Lightcurves for these objects are shown in Figure \ref{lc_SU}.  These stars have been identified as VW Hydri, a subclass of SU-UMa Cataclysmic Variables, itself a subclass of dwarf novae distinguished by superhumps \citep{Cannizzo2012,Kato2004}. These stars, and the dwarf nova class of stars in general, have been the subject of extensive study for the past century as these semi-detached binary systems present unique, periodic outburst behaviors. The first reference to V1504 Cyg comes from \citet{Kukarkin1977} and discovery of V344 Lyr from \citet{Hoffmeister1966}. Both appear in the original \textit{A catalog and Atlas of Cataclysmic Variables}  by \citet{Downes1993}. \citet{Cannizzo2012} present a study of the outburst properties of these two CVs contained in the Kepler field.

\subsubsection{Rare Objects}\label{sec_rarities}
\begin{figure}
\gridline{\fig{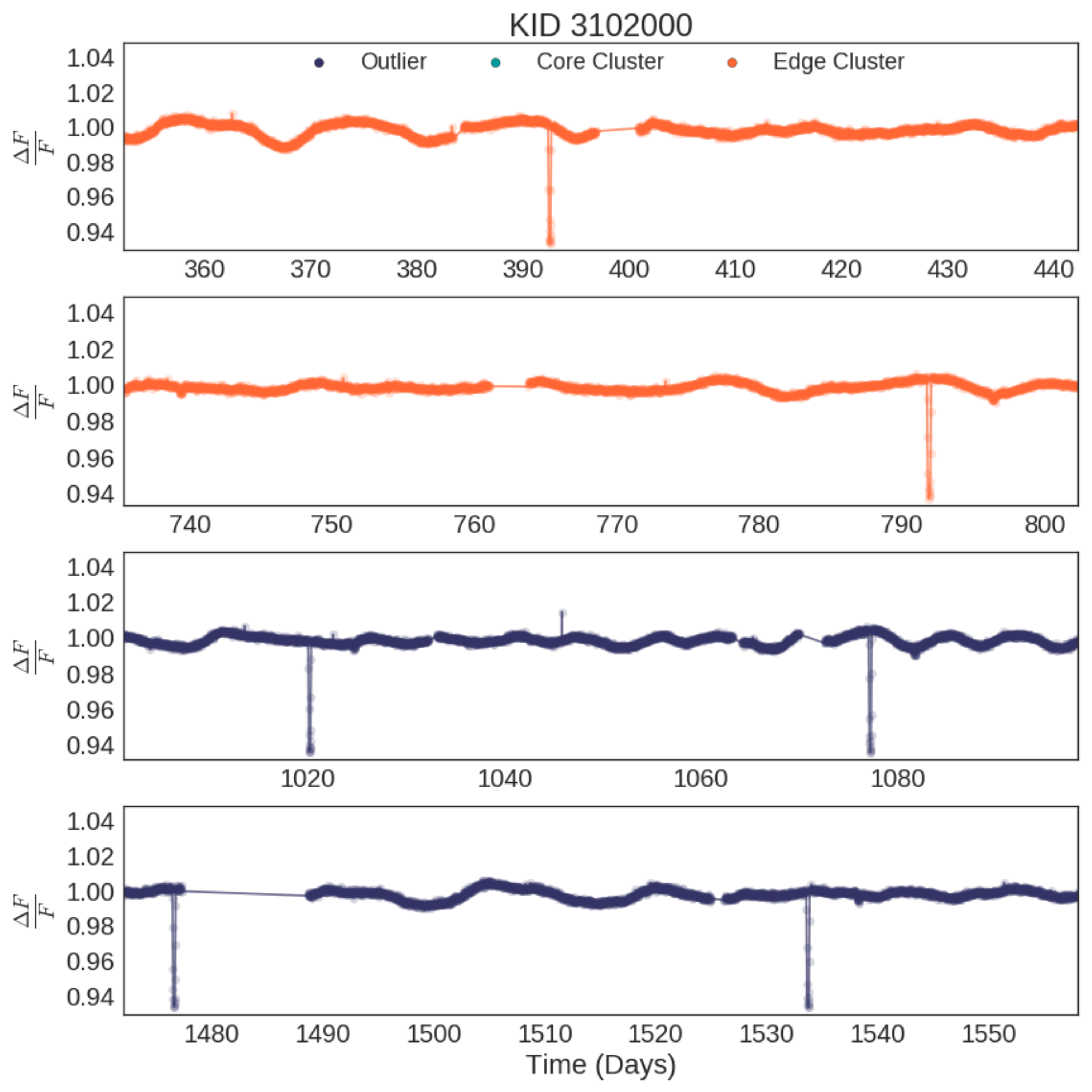}{0.5\textwidth}{(a) \footnotesize KIC 3102000 is identified as a longterm eclipsing binary and as an outlier in Quarters 11 and 16.}
\fig{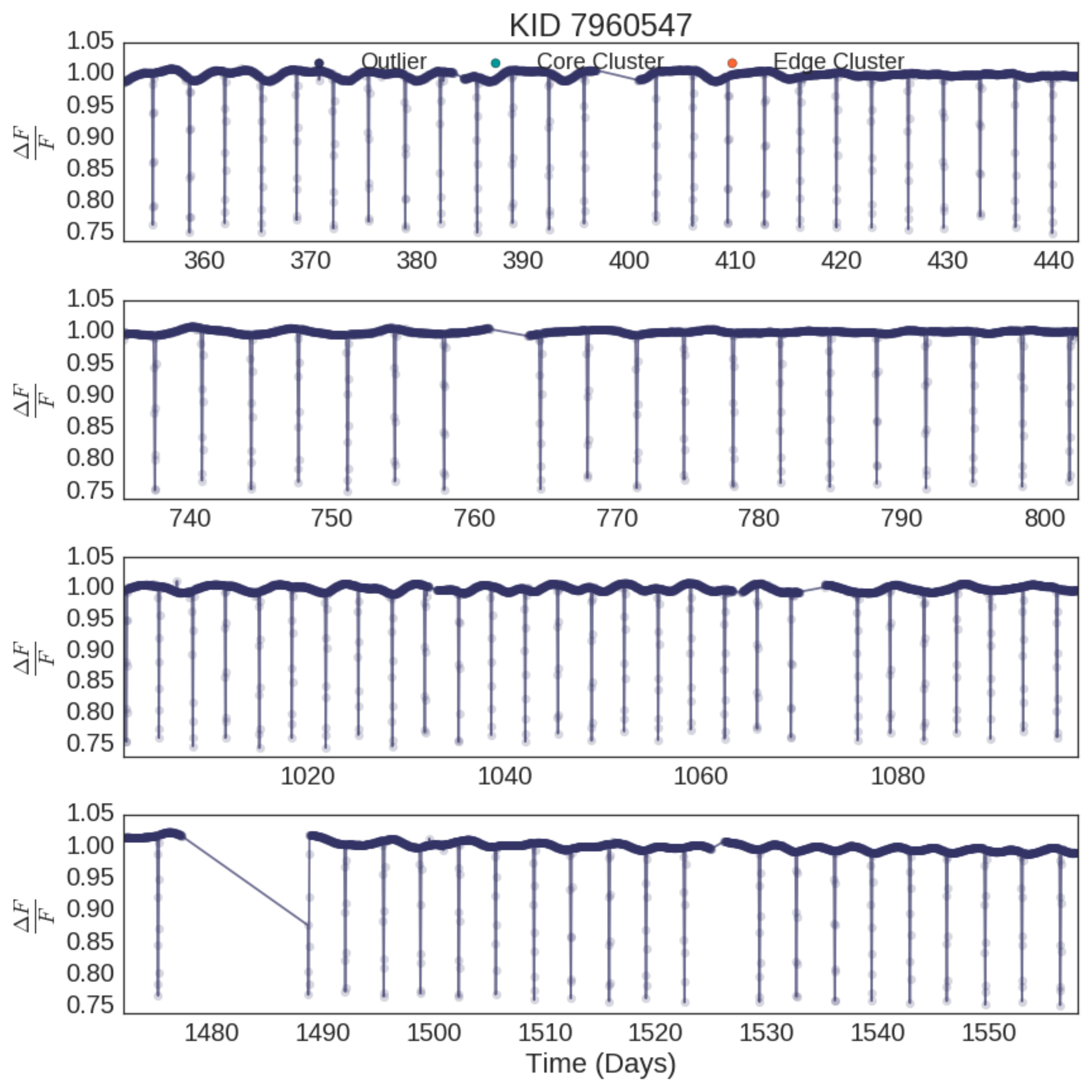}{0.5\textwidth}{(b) \footnotesize KIC 7960547 shows standard behaviour for an eclipsing binary.}}
\gridline{\fig{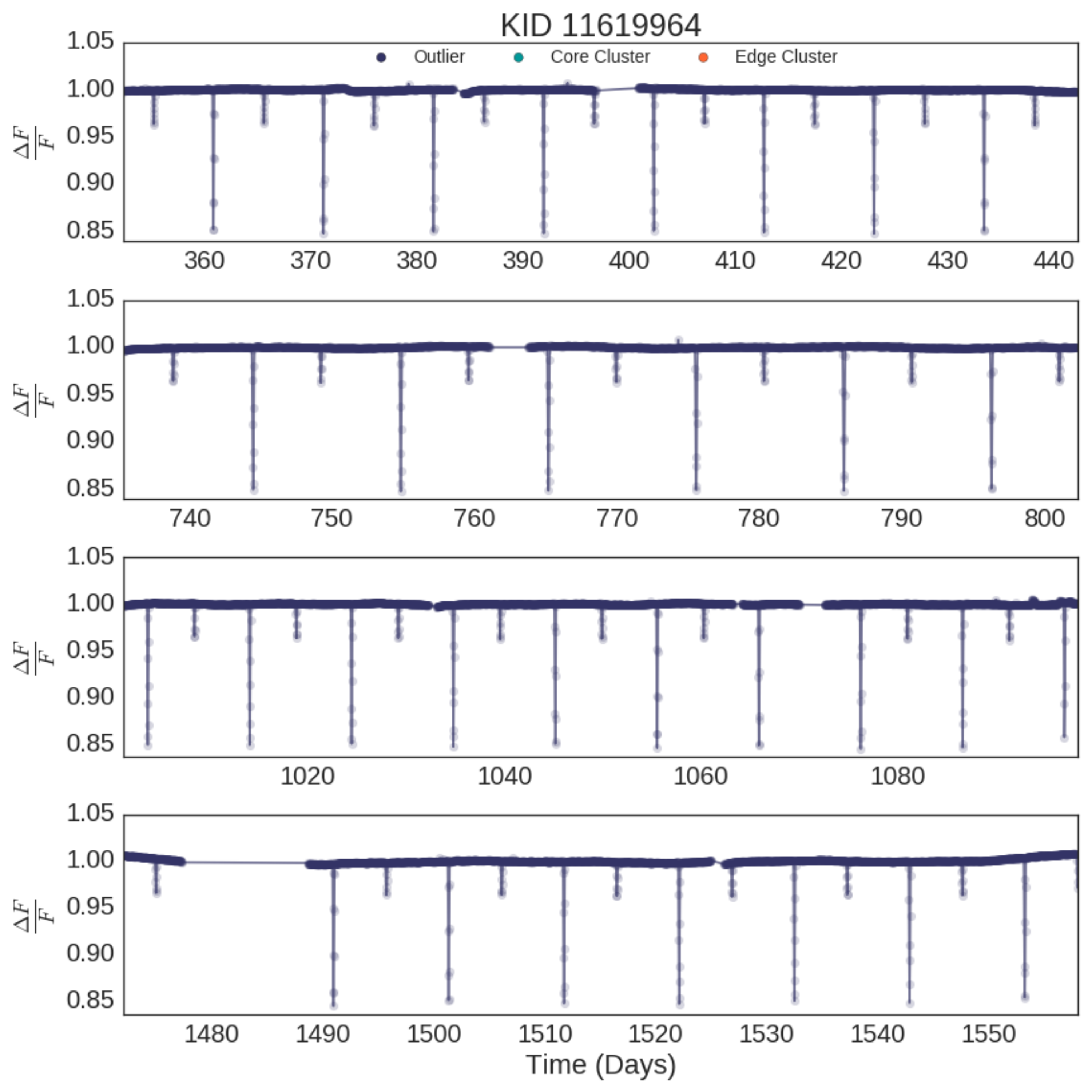}{0.5\textwidth}{(c) \footnotesize KIC 11619964 exhibits standard behavior for an eclipsing binary with regular transits of two unique depths.}
\fig{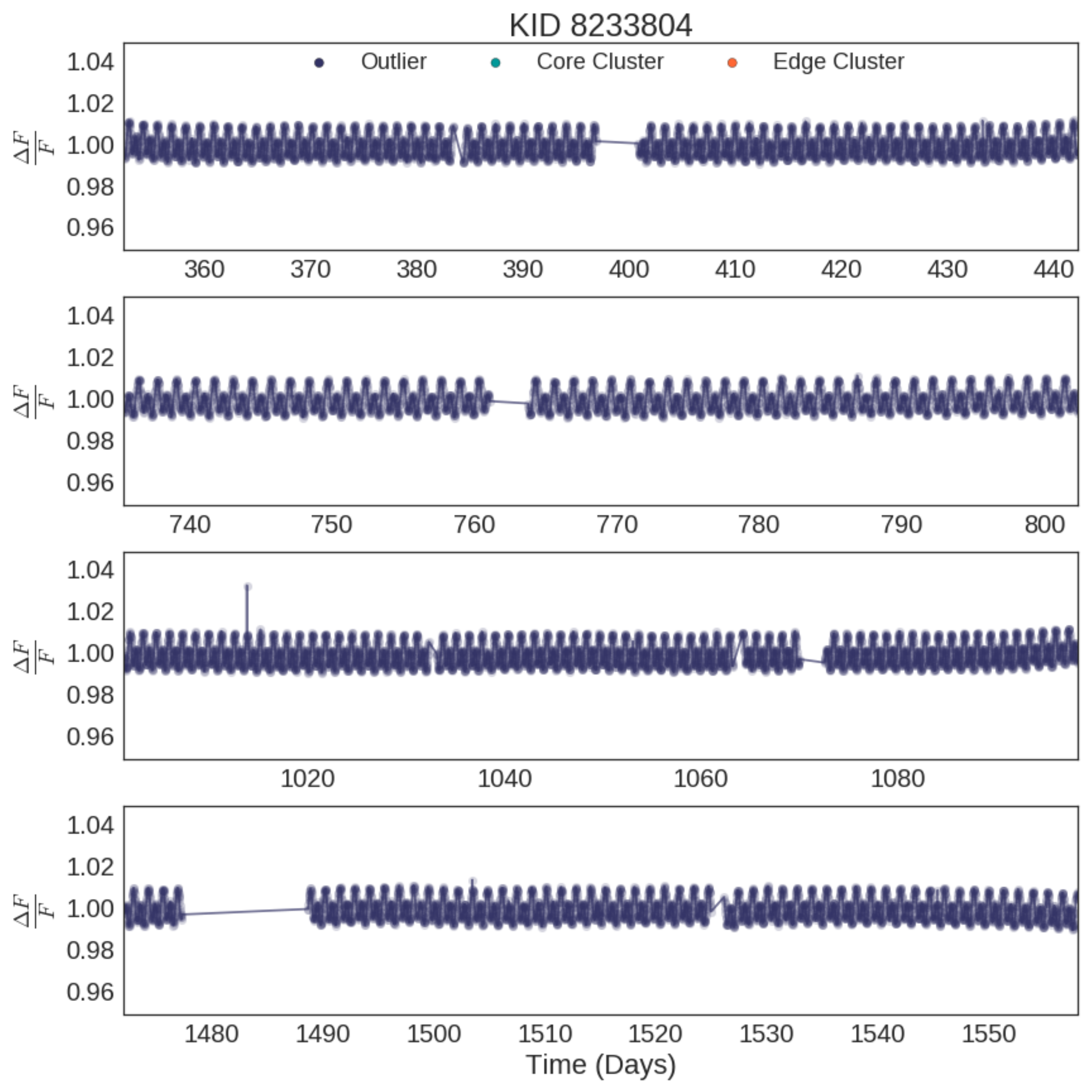}{0.5\textwidth}{(d) \footnotesize KIC 8233804 B type star. Its variability is likely due to rotation or binarity\citep{McNamara2012}.}}\caption{Objects that are relatively rare within the Kepler dataset, like eclipsing binaries in (a), (b), and (c) and the B-type star in (d), are determined to by outliers by our method.}\label{lcs_rare}
\end{figure}

As we search for anomalies, we are, by definition, searching for rare objects. Rarity is determined by the data each object is compared to. When looking at a subset of data, the rarity of different objects may differ from their rarity in the full set.
Three of the outliers from our example objects have been identified as eclipsing binary stars, 
KID 3102000, 
KID 7960547, and 
KID 11619964. 
Notably, none of these three objects were found to be core cluster members in any quarter. This would suggest that binary stars are relatively rare in the Kepler data set. The Kepler Eclipsing Binary Catalog contains 2878 entries at the time of this writing, accounting for only 1.3\% of Kepler objects \citep{Kirk2016}.
KIC 3102000 is a highly eccentric, long period, detached eclipsing binary system of spectral type G1V (T\textsubscript{eff}=5933K), with a period of 57.06 days. It was identified as a Kepler Object of Interest with clear transit activity by \citet{Tenenbaum2012}, but \citet{Dong2013} later recognized that it was in fact an eclipsing binary with a highly eccentric (e\textsubscript{min}=0.73) period. The lightcurve for KIC 3102000 is shown in Figure \ref{lcs_rare}(a).

KIC 7960547 was identified as a detached eclipsing binary with a transit period of 6.767 days by \citet{Prsa2011} in the first comprehensive catalog of eclipsing binaries for the Kepler field and as a KOI based on transit activity (T\textsubscript{eff}=5669K, T\textsubscript{1}/T\textsubscript{2}=0.94428, R\textsubscript{1}+R\textsubscript{2}=0.07233R\textsubscript{\(\odot\)}, sin i=1.00059, Fig. \ref{lcs_rare}(b)).

KIC 11619964 was identified as a detached eclipsing binary with a transit period of 10.368 days by \citet{Prsa2011} in the first comprehensive catalog of eclipsing binaries for the Kepler field and had properties refined by \citet{Slawson2011} (T\textsubscript{eff}=5582K, log g=4.42, T\textsubscript{1}/T\textsubscript{2}=0.87462, R\textsubscript{1}+R\textsubscript{2}=0.09842\textsubscript{\(\odot\)}, sin i=0.99486, Fig. \ref{lcs_rare}(c)).

KIC 8233804 is a main sequence B9V star(T\textsubscript{eff}=11068K, log g=4.377,mass=2.102M\textsubscript{\(\odot\)}, radius=1.555R\textsubscript{\(\odot\)}). As discussed in Section \ref{sec_keplerdata}, the Kepler Input Catalog is curated to find exoplanets and there are very few B-type stars. \citet{McNamara2012} studied 252 B-star candidates and classified this object's variability due to rotation or binarity rather than pulsations given its relative smoothness. The lightcurve for KIC 8233809 is shown in Figure \ref{lcs_rare}(d).

\subsubsection{Edge of cluster outliers}\label{sec_edges}
\begin{figure} \centering
\gridline{
\fig{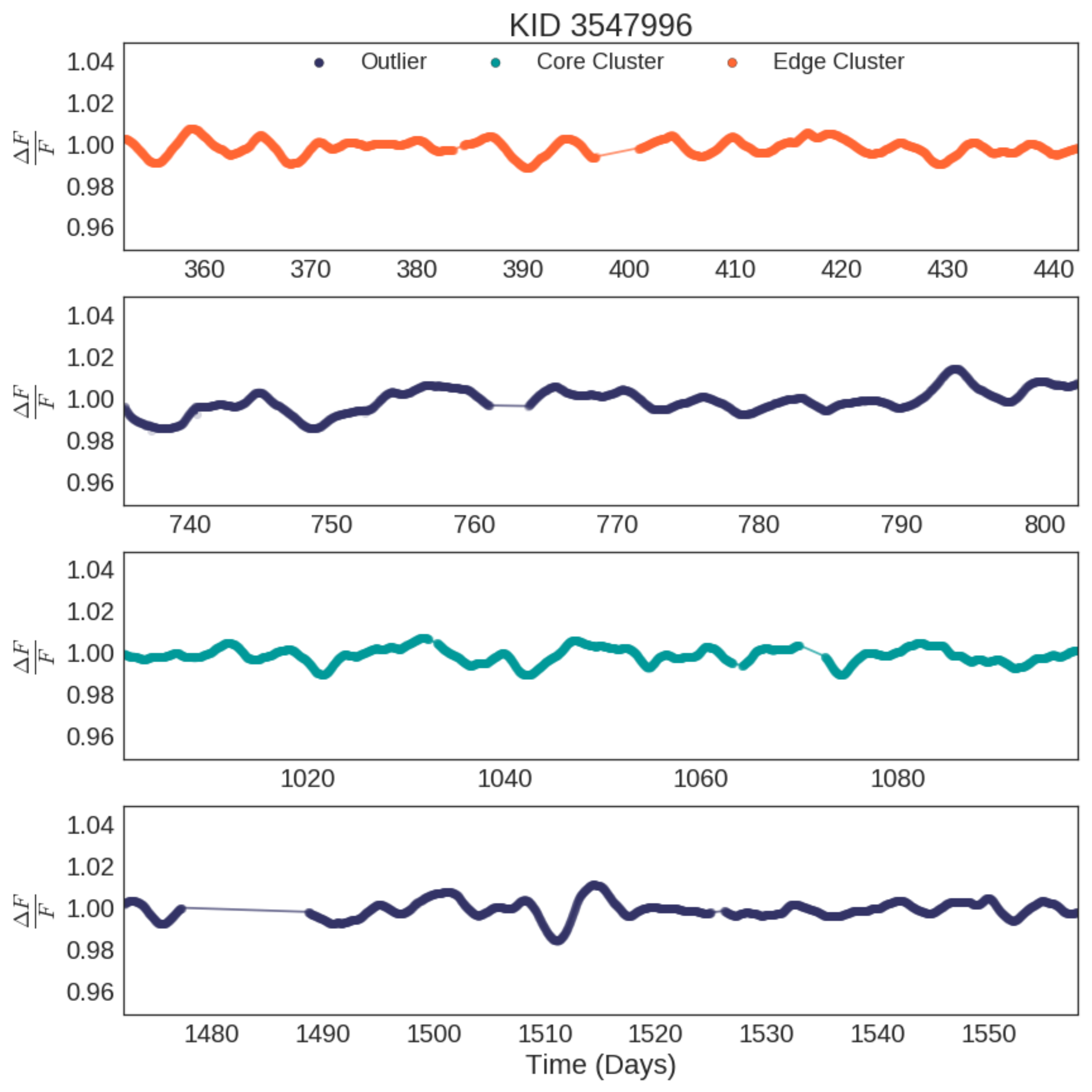}{0.5\textwidth}{(a) \footnotesize KIC 3547996 is identified as an outlier in Quarters 8 and 16. This object has been identified as a star by the GAIA \cite{Gaia2016}.}
\fig{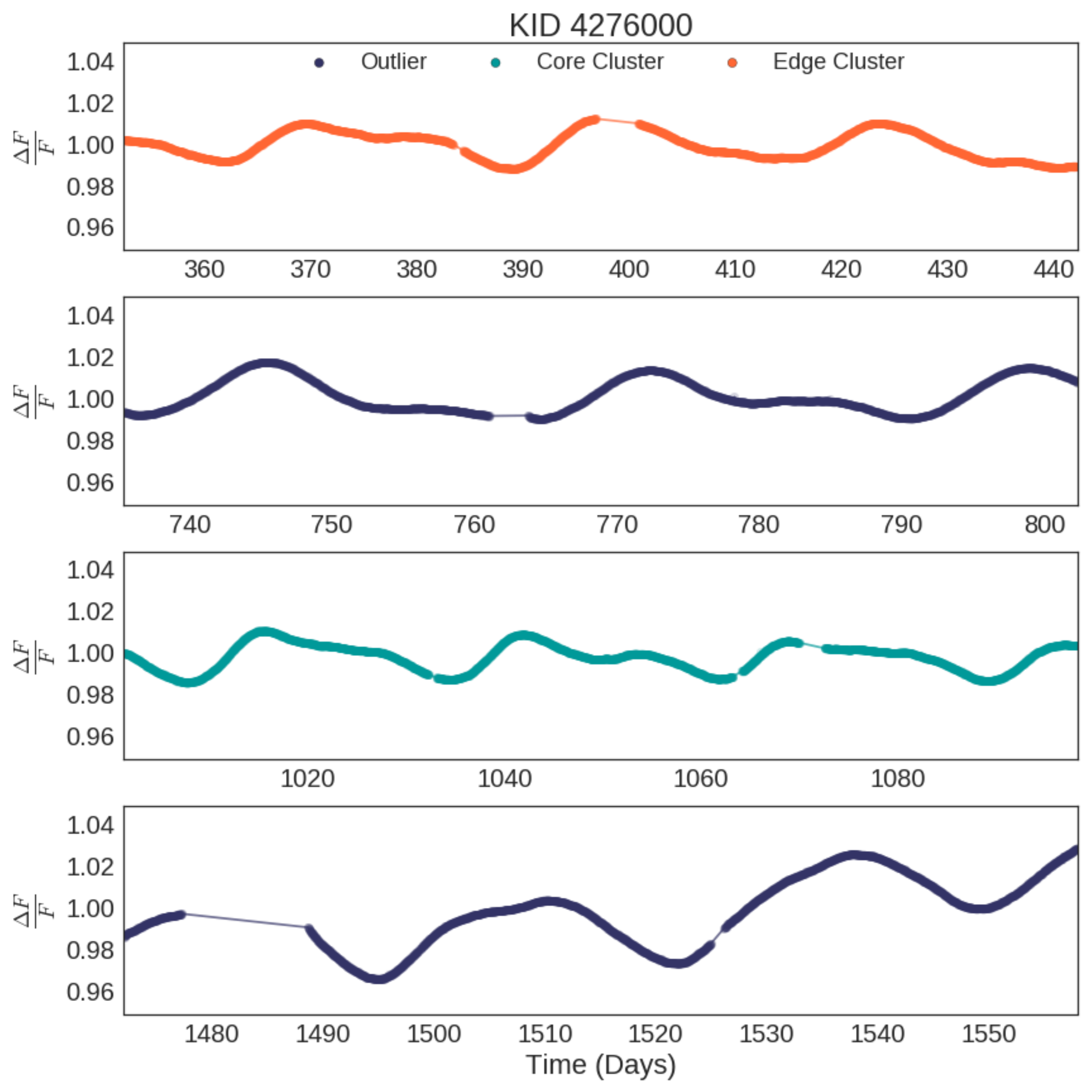}{0.5\textwidth}{(b) \footnotesize KIC 4276000 moves between all identifications and appears as an outlier twice in Quarters 8 and 16. It has been identified as a rotationally variable star by \citet{Debosscher2011}.}
}\caption{\footnotesize Some objects are identified as outliers, core cluster members, and edge cluster members. An examination of the reductions in \ref{all_annotated} show that these objects maintain similar relationships relative to the rest of the population in each quarter, consistently on the edge, fluctuating between outlier and cluster member.}\label{lcs_edge}
\end{figure}

There is a particular challenge in clustering associated with sparsity at cluster edges that makes the distinction between clustered data and unclustered data nebulous.
In Gaussian distributions, a majority of data lies near the mean with sparser data in the tails. Any cluster with one or more Gaussian distributed features will inevitably have an uneven density with sparser data at the edges. Two objects, KID 3547996 (Figure \ref{lcs_edge}(a)) and KID 4276000 (Figure \ref{lcs_edge}(b)), are sometimes identified as outliers, and as edge cluster members at other times. These two objects exist on the edge of the clustered data, oscillating from quarter to quarter in relation to other data. Rather than being truly anomalous, these would appear to vary in designation due to the relative sparsity at the edge of the cluster. 

KIC 3547996 is contained in the Simbad database as TYC 3134-24-1 from the Tycho-2 Catalog of the brightest 2.5 million stars and as 2MASS J19291519+3841290. This star appears to be a K7 III giant (T\textsubscript{eff}=4023K, log g=1.091, mass=1.63M\textsubscript{\(\odot\)}, radius 60.23R\textsubscript{\(\odot\)}). The lightcurve for KIC 3547996 is shown in Figure \ref{lcs_edge}(a).

KIC 4276000 appears to be a G6 IV subgiant (T\textsubscript{eff}=5510K, log g=3.508, mass=1.79M\textsubscript{\(\odot\)}, radius= 3.901R\textsubscript{\(\odot\)}). This object was classified by \citet{Debosscher2011} as a rotationally variable star following the Q0 and Q1 data release via an automated method, a new classification made by this paper unique from other sources of stellar variability. The lightcurve for KIC 4276000 is shown in Figure \ref{lcs_edge}(b).

\subsubsection{Data Artifacts}\label{artifacts}
\begin{figure}
\gridline{
\fig{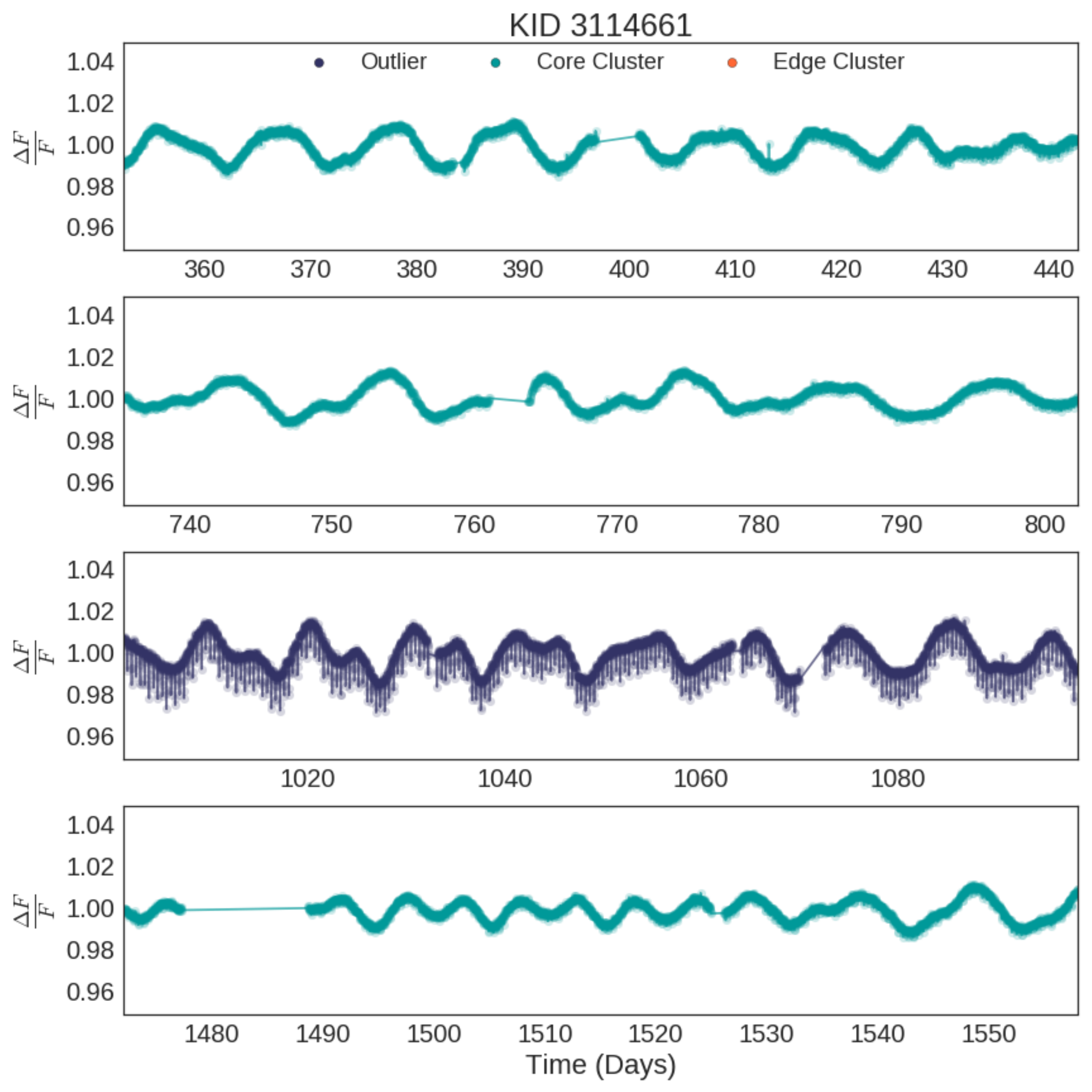}{0.5\textwidth}{(a) KIC 3114661 shows exaggerated features in Quarter 11, apparently related to the quarterly roll that Kepler performed.}
\fig{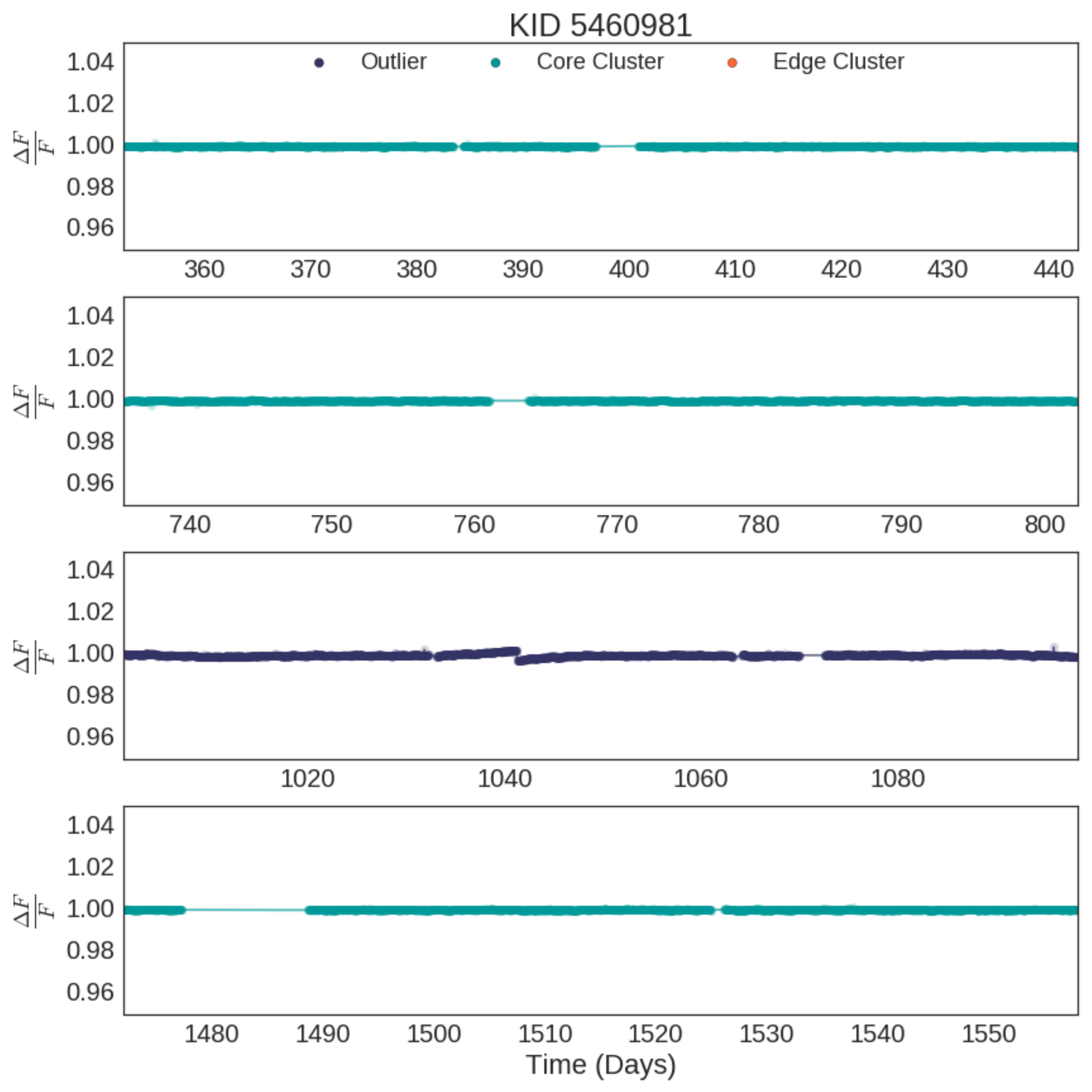}{0.5\textwidth}{(b) KIC 5460981 is identified as an outlier only in Quarter 11 where there is odd jump around day 1040.}
}
\gridline{
\fig{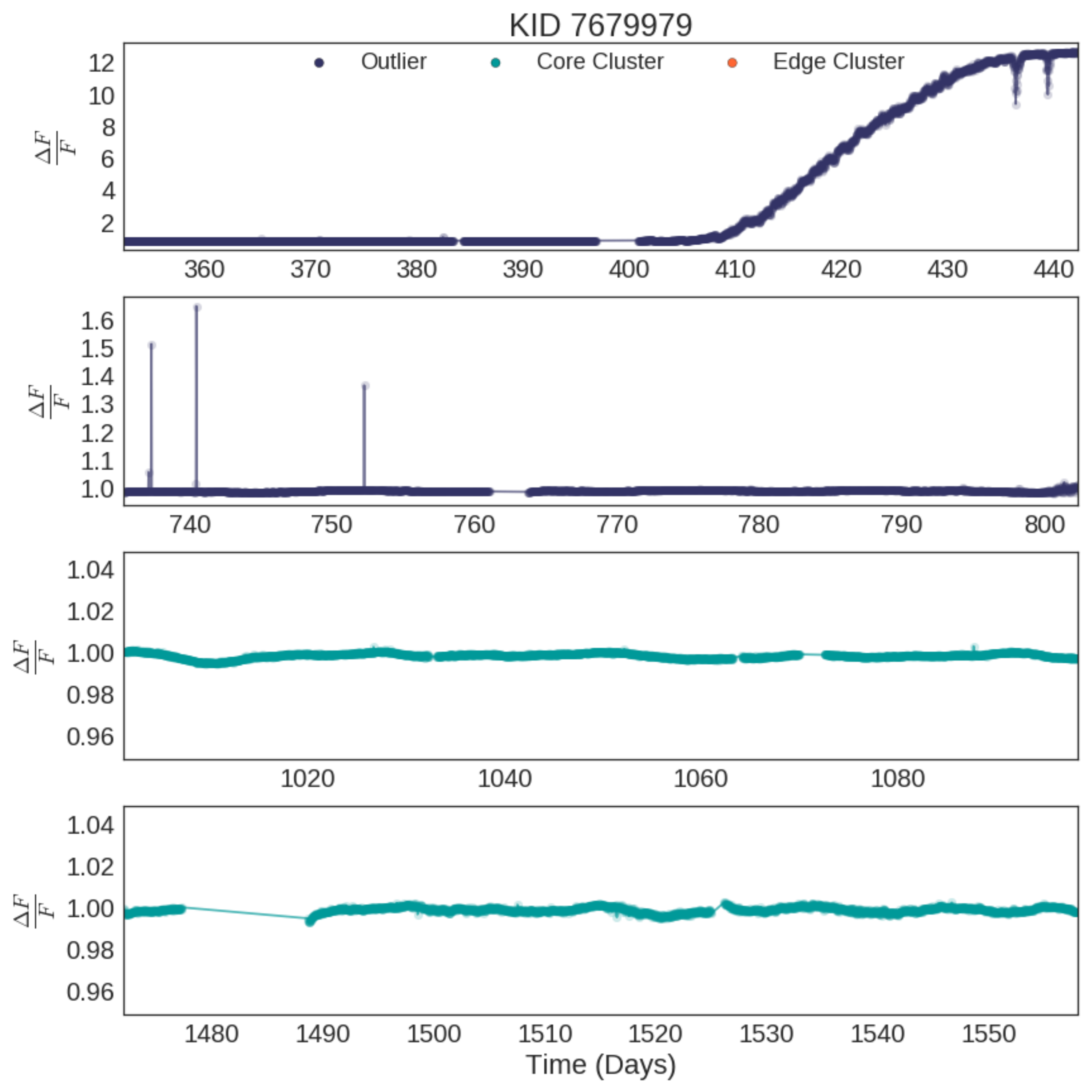}{0.45\textwidth}{(c) KIC 7679979 exhibits odd behaviours in Quarters 4 and 8 that are not present in any other quarter.}
\fig{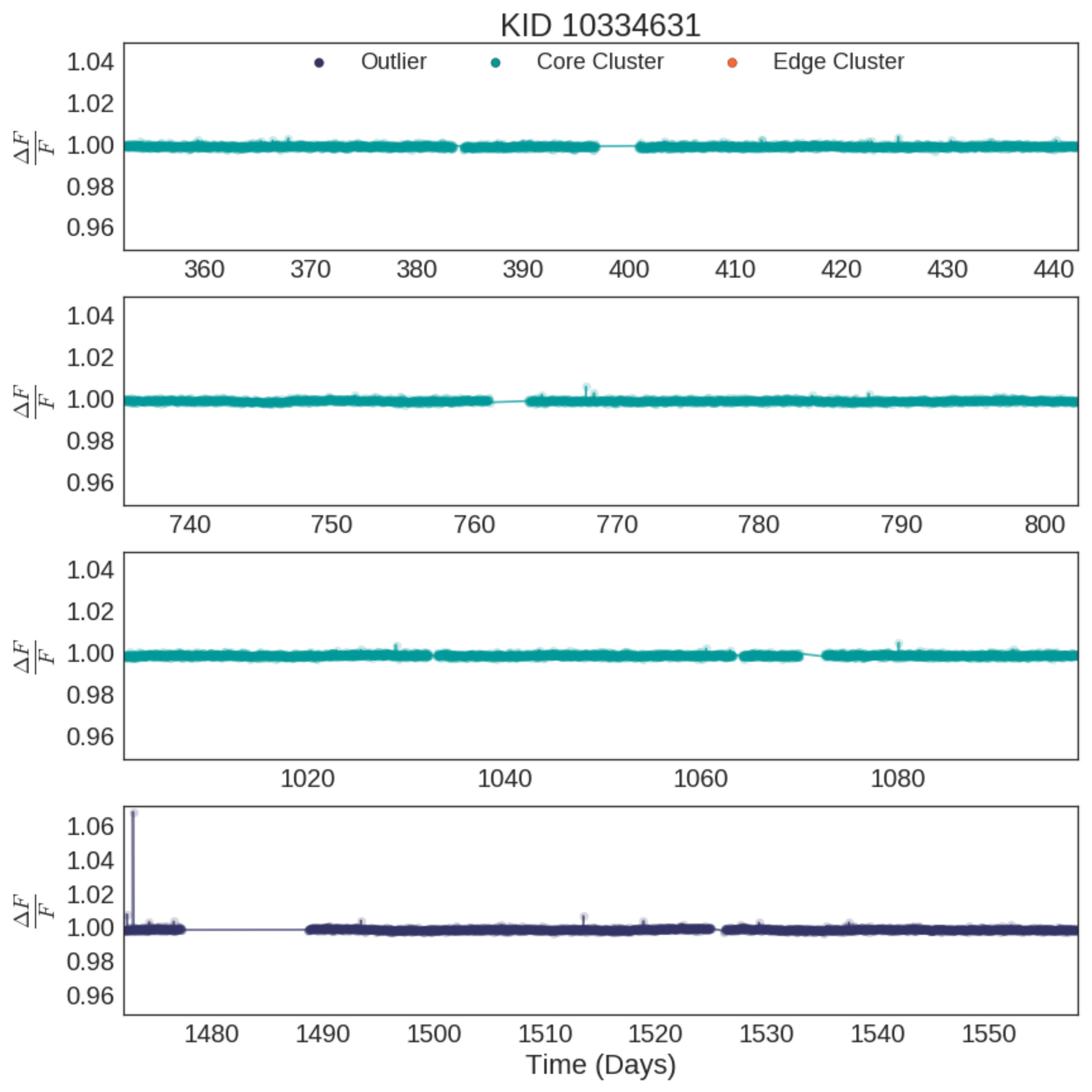}{0.45\textwidth}{(d) KIC10334631 has an odd jump at the beginning of Quarter 16, where it is identified as an outlier.}
}\caption{Many of the outliers our method identified exhibit features that appear to be artifacts from data processing or collection, rather than behaviors of the objects themselves. In either case, our method is able to identify objects with anomalous behavior.}\label{lcs_artifacts}
\end{figure}

In addition to finding novel or rare phenomena, anomaly detection can be useful in finding aberrant data caused by processing. This can help refine data processing techniques, clean databases of corrupt data, and reveal unaccounted for systematics.
Several of the identified outliers, KID 3114661 (Fig. \ref{lcs_artifacts}(a)), KID 5460981 (Fig. \ref{lcs_artifacts}(b)), KID 7679979 (Fig. \ref{lcs_artifacts}(c)), and KID 10334631 (Fig. \ref{lcs_artifacts}(d)), are core cluster members except for quarters where features would appear to be related to data artifacts rather than the behavior of the objects themselves.

KIC 3114661 appears to be a G7 VI subdwarf (T\textsubscript{eff}=5243K, log g=4.674, mass=0.69M\textsubscript{\(\odot\)}, radius=0.635R\textsubscript{\(\odot\)}). It was identified in a study by \citet{Walkowicz2013} as a rotationally variable star and as a potential planet candidate. This was identified later by \citet{Coughlin2014} to be a false-positive via ephemeris through direct pixel response function and by \citet{Morton2016} by calculation of the False Positive Probability. KID 3114661 exhibits a persistent and periodic variance which is exaggerated in Quarter 11. Examining other quarters of this objects data reveals that this behavior coincides neatly with the quarterly roll that Kepler performed. The lightcurve for KIC 3114661 is shown in Figure \ref{lcs_artifacts}(a).

KIC 5460981 is a main sequence F5 V star (T\textsubscript{eff}=6692K, log g=4.275, mass=1.26M\textsubscript{\(\odot\)}, radius=1.351R\textsubscript{\(\odot\)}). \citet{McQuillan2014} determined this object to have no significant period detection or transit features. The lightcurve of KID 5460981 shows an odd jump around day 1040. The lightcurve for KIC 5460981 is shown in Figure \ref{lcs_artifacts}(b).

KIC 7679979 appears to be a main sequence G3VI subdwarf (T\textsubscript{eff}=5693K, log g=4.551, mass=0.95M\textsubscript{\(\odot\)} and radius 0.856R\textsubscript{\(\odot\)}) with no previously documented variability or properties of note. The lightcurve of KID7679979 exhibits abnormal brightening at the end of Quarter 4 and shows some odd spikes in Quarter 11 around days 740 and 750, as can be seen in Figure \ref{lcs_artifacts}(c). This object is identified only as a star by \cite{McQuillan2014}. This does not appear to coincide with any known phenomena nor do either of these behaviors present themselves in other quarters.

KIC 10334631 appears to be a K1VI subdwarf (T\textsubscript{eff}=4991K, log g=4.631, mass=0.852M\textsubscript{\(\odot\)}, radius=0.944R\textsubscript{\(\odot\)}. KID10334631 exhibits an individual, bright observation exceeding a 6\% increase around day 1470. The lightcurve for KIC 10334631 is shown in Figure \ref{lcs_artifacts}(d).

\subsection{Discussion}\label{sec_discussion}
These example outliers show that our method is adept at identifying anomalous behavior where it occurs including rare objects within a dataset and data potentially corrupted. Notably, however, not every outlier identified exhibited behavior we might recognize as anomalous. Objects on the edge of clusters that do not exhibit particularly interesting behavior, are identified as outlying due to the non-uniform density of data given the features we consider. This poses a problem for our method, as we define a cluster based on its density. However, the consistency of relative position from quarter to quarter for such objects may present an opportunity to utilize intra-quarter movement to further characterize an objects outlying nature.
This method effectively pares datasets down to the most promising potential objects containing anomalous behavior. It does not, however, have the ability to identify the cause of this behavior or the ability to predict what objects might be of interest. The use of this method is to supplement surveys by focusing the search for novel phenomena and data anomalies to a manageable size, so that as data is made available, additional observation may be focused more effectively.

\section{Conclusions \& Future Work}\label{sec_conclusion}
We have demonstrated the effectiveness of our method on Kepler data by successfully identifying anomalous behavior of Boyajian's star in addition to showing how this method can quickly identify interesting subsamples (like eclipsing binaries), truly unique or rare objects (cataclysmic variables, Boyajian's star, or data artifacts).  We have also seen that in an individual quarter, the list of outliers will contain objects that consistently reside on or near the edge of clustered data, but which do not present anomalous behavior. 
This demonstrates a limitation of the methodology, clusters are defined to have constant density, but evidently do not. Gaussian distributions within clustered data lead to sparser regions at the edges of clusters. Problematically, this also appears to be where some of the genuinely anomalous data resides. We see, though, that a potential solution presents itself if we examine intra-quarter movement.

Following this work, we will examine outlier distributions in more depth with particular attention paid to formalize outlier scoring. Having established the utility of this method using a feature set curated for classification and datamining, we will evaluate the utility of features in the interest of maximizing impact while reducing computational cost. We will examine the utility of other outlier identification methods as well, and apply these methods to the full Kepler dataset of long cadence lightcurves. We also intend to apply our method to other time-series data as they become available, including the Transiting Exoplanet Survey Satellite (TESS), and eventually LSST. As larger scale surveys release data, computational methods will be relied upon to identify novel sources. \citet{Wagstaff2013} identify the need to not only develop methods of outlier identification and rank, but further to develop a diversity of methods to highlight outliers of different types and of different origins as we seek to enable scientific discovery through data prioritization. 

\section{Acknowledgments}
DG acknowledges the support of the Illinois Space Grant Consortium Graduate Fellowship and thanks the LSSTC Data Science Fellowship Program, his time as a Fellow has benefited this work. LW thanks the New Frontiers in Astronomy and Cosmology grant, administered by Don York of the University of Chicago and funded by the John Templeton Foundation, for support of initial development of the methods used herein, and collaborators Revant Nayar, Ed Turner, Jeff Scargle, Vikki Meadows, and Tony Zee for their contributions to this project's inception.
All of the data presented in this paper includes data collected by the Kepler mission and were obtained from the Mikulski Archive for Space Telescopes (MAST). STScI is operated by the Association of Universities for Research in Astronomy, Inc., under NASA contract NAS5-26555. Funding for the Kepler mission is provided by the NASA Science Mission directorate.
Stellar properties reported are the revised stellar properties of Kepler Targets reported by \citet{Huber2014} except where otherwise noted, accessed through the NASA Exoplanet Archive, operated by the California Institute of Technology under contract with the National Aeronautics and Space Administration under the Exoplanet Exploration Program.
This research has made use of the SIMBAD database, operated at CDS, Strasbourg, France \citep{Wenger2000}.

\software{Scikit-learn \citep{scikit-learn}, Astropy \citep{astropy}, SciPy \citep{scipy}, NumPy \citep{numpy}, Matplotlib \citep{matplotlib}, Pandas \citep{pandas}}
\newpage
\appendix
\section{Features}\label{app_features}
\startlongtable
%\begin{deluxetable}{lll}
\begin{deluxetable}{p{.35\textwidth}lp{.5\textwidth}}
\tablecolumns{3}
\tablewidth{1.0\columnwidth}
%% This is the title of the table.
\tablecaption{Features}

%\tablehead{\multicolumn{1}{l}{Feature} & \multicolumn{1}{l}{Code reference} & \multicolumn{1}{l}{Description}} 
\tablehead{Feature & Code Reference & Description}
%% All data must appear between the \startdata and \enddata commands
\startdata
Long-term trend   & longtermtrend   &  Linear trend of fluxes over whole series \\
Mean to median ratio   & meanmedrat   &  Ratio between the mean and the median \\
Skewness of fluxes  & skews   &  Skewness of the fluxes \\
Variance  & varss   &  Variance of the fluxes \\
Coefficient of variability   & coeffvar   &  Ratio of the standard deviation to the mean of the fluxes  \\
Standard deviation & stds   &  Standard deviation of the fluxes \\
Number of outliers beyond 1$\sigma$   & numout1s   &   Count of flux values beyond 1$\sigma$ deviation from the mean \\
Number of negative outliers   & numnegoutliers   &   Count of flux values beyond 4$\sigma$ deviation from the mean, less than the mean  \\
Number of positive outliers   & numposoutliers   &   Count of flux values beyond 4$\sigma$ deviation from the mean, greater than the mean  \\
Number of outliers   & numoutliers   &   Count of total flux values beyond 4$\sigma$ deviation from the mean  \\
Kurtosis   & kurt   &   Kurtosis of the fluxes\\
Median Absolute Difference   & mad   &   Median absolute difference from the median flux  \\
Maximum slopes   & maxslope   &   Value defining the 99th percentile of slopes between 2 sequential fluxes\\
Minimum slopes   & minslope   &   Value defining the 1st percentile of slopes between 2 sequential fluxes\\
Mean of postive slopes  & meanpslope   & Mean of positive slopes between 2 sequential fluxes \\
Mean of negative slopes  & meannslope   & Mean of negative slopes between 2 sequential fluxes \\
G assymetry   & g\_asymm   &   Ratio of mean positive slopes to negative slopes (large dummy value of 10 given if no negative slopes)  \\
Rough g assymetry   & rough\_g\_asymm   &   Ratio of number of positive slopes to negative slopes (large dummy value of 10 given if no negative slopes)  \\
Diffence assymetry   & diff\_asymm   &   Difference between the mean of the positive slopes and the absolute mean of the negative slopes  \\
Skewness of slopes  & skewslope   &  Skewness of slopes\\
Slopes mean value  & meanabsslope   & Mean of the absolute value slopes \\
Variance of absolute slopes & varabsslope   &  Variance of the absolute value slopes\\
Variance of the slopes  & varslope   &  Variance of the slopes\\
Absolute mean of the \phm{secondsecond} second derivative  & absmeansecder   & Mean of the absolute second derivative of the fluxes \\
Number of positive spikes   & num\_pspikes   &   Count of positive spikes as defined by a positive slope 3$\sigma$ or greater than the mean positive slope.  \\
Number of negative spikes   & num\_nspikes   &   Count of negative spikes as defined by a negative slope 3$\sigma$ or smaller than the mean negative slope.  \\
Number of positive second derivative spikes   & num\_psdspikes   &   Count of second derivative values beyond 4$\sigma$ deviation from the mean of second derivative values, greater than the mean  \\
Number of negative second derivative spikes   & num\_nsdspikes   &   Count of second derivative values beyond 4$\sigma$ deviation from the mean of second derivative values, less than the mean \\
Standard deviation ratio   & stdratio   &   Ratio of the standard deviation of the positive slopes to the standard deviation of the negative slopes (large dummy value of 10 given if negative standard deviation is zero)  \\
Pair slope trend   & pstrend   &   Ratio of positive slopes with a subsequent positive slope to the total number of slopes  \\
Number of 'zero' crossings   & num\_zcross   &   Count of occurences where sequential observations cross the longterm trendline \\
Number of 'plus-minus' slope switches   & num\_pm   &   Count of slope transitions from positive to negative  \\
Length of naive maxima   & len\_nmax   &   Count of naive maxima where a maxima is the largest flux value within 10 points on either side  \\
Length of naive minima   & len\_nmin   &   Count of naive minima where a minima is the smallest flux value within 10 points on either side  \\
Maxima auto-correlation coefficient   & mautocorrcoef   &   Auto-correlation coefficient of one maxima to the next \\
Peak-to-peak slopes   & ptpslopes   &   Mean of the slopes from naive peak-to-peak  \\
Periodicity   & periodicity   &   Coefficient of variability for time-differences, ratio of the standard deviation to the mean of time-difference between maxima  \\
Periodicity residual   & periodicityr   &   Coefficient of variability for time-differences of maxima using residuals  \\
Naive periodicity   & naiveperiod   &   Mean of the time-differences between naive maxima  \\
Maxima variation   & maxvars   &   Coefficient of variation of the maxima, ratio of the standard deviation to the mean of the naive maxima flux values  \\
Maxima variation residuals   & maxvarsr   &   Coefficient of variation of maxima flux values using residuals instead of standard deviation  \\
Odd to even ratio   & oeratio   &   Ratio of odd indice minima flux values to even indice means flux values  \\
Amplitude analogue   & amp\_2   &   Peak-to-peak based on 1st and 99th percentile  \\
Normalized amplitude analogue   & normamp   &   amp\_2 divided by the mean flux value  \\
Median buffer percentile   & mbp   &   Fraction of the number of points within 10\% of the amplitude to the median  \\
Ratio of flux percentiles \phm{(dthtonth)} (60th to 40th)  & mid20   &  Ratio of flux percentiles \phm{stringstringstringstringstring} (60th to 40th) over (95th to 5th)\\
Ratio of flux percentiles \phm{(dthtonth)} (67th to 32nd)  & mid35   &  Ratio of flux percentiles \phm{stringstringstringstringstring} (67th to 32nd) over (95th to 5th)\\
Ratio of flux percentiles \phm{(dthtonth)} (75th to 25th)  & mid50   &  Ratio of flux percentiles \phm{stringstringstringstringstring} (75th to 25th) over (95th to 5th)\\
Ratio of flux percentiles \phm{(dthtonth)} (82nd to 17th)  & mid65   &  Ratio of flux percentiles \phm{stringstringstringstringstring} (82nd to 17th) over (95th to 5th)\\
Ratio of flux percentiles \phm{(dthtonth)} (90th to 10th)  & mid80   &  Ratio of flux percentiles \phm{stringstringstringstringstring} (90th to 10th) over (95th to 5th)\\
Percent amplitude   & percentamp   &   Largest absolute difference between the max or min flux and the median (as a percentage of the median)  \\
Maximum ratio   & magratio   &   Ratio of the maximum flux value to amp\_2  \\
Auto-correletion coefficient   & autocorrcoef   &   Auto-correlation coefficient of the fluxes  \\
Slopes auto-correlation coefficient   & sautocorrcoef   &   Auto-correlation coefficient of the slopes  \\
Flatness mean around naive maxima   & flatmean   &   Mean average of flatness values around maxima, where flatness is the mean of the absolute value of 6 slopes on either side of each maxima  \\
Flatness mean around naive minima   & tflatmean   &   Mean average of flatness values around minima, where flatness is the mean of the absolute value of 6 slopes on either side of each minima  \\
Roundness mean around \phm{stringstring} naive maxima   & roundmean   &   Mean average of roundness values around maxima, where roundness is the mean of 6 second derivative values on either side of maxima  \\
Roundness mean around \phm{stringstring} naive minima   & troundmean   &   Mean average of roundness values around minima, where roundness is the mean of 6 second derivative values on either side of minima  \\
Roundness ratio   & roundrat   &   Ratio of roundness of maxima to roundness of minima  \\
Flatness ratio   & flatrat   &   Ratio of flatness of maxima to flatness of minima  \\
\enddata
\tablecomments{Sample code snippets for all features are available at \\https://github.com/d-giles/KeplerML/blob/master/feature\_key.txt}
\end{deluxetable}

\newpage
\bibliographystyle{aasjournal}
\bibliography{bibliography}
\end{document}